\documentclass{article}
\usepackage[a4paper,top=3cm,bottom=3.5cm,left=3.2cm,right=3.2cm]{geometry}

\usepackage[utf8]{inputenc}
\usepackage{authblk}
\usepackage{graphicx}
\usepackage{subfigure}
\usepackage{amssymb}
\usepackage{lineno}
\usepackage[english]{babel}
\usepackage{palatino, url}
\usepackage{amsmath}
\usepackage{hyperref}

\usepackage{listings}
\usepackage{graphicx}
\usepackage[rightcaption]{sidecap}
\usepackage{textcomp}
\usepackage{array} 
\usepackage{booktabs} 
\usepackage{multirow} 
\usepackage{caption} 
\usepackage{floatflt}
\usepackage{rotating} 
\usepackage{scrextend}
\usepackage{multirow}
\usepackage[numbers]{natbib}
\usepackage{soul}
\usepackage{mathtools}

\title{Critical slowing down in hive temperature does not provide actionable early warnings of honey bee colony collapse}

\author[1]{Daniele Proverbio}
\author[2]{Els Weinans}
\author[3]{Davide Frizzera}
\author[3]{Desiderato Annoscia}
\author[3]{Francesco Nazzi}
\author[1]{Giulia Giordano}
\affil[1]{\small Department of Industrial Engineering, University of Trento, Trento 38123, IT}
\affil[2]{\small Copernicus Institute of Sustainable Development, Utrecht University, 3584 CB Utrecht, NL}
\affil[3]{\small Department of Agricultural, Food, Environmental and Animal Sciences, University of Udine, Via delle Scienze 206, 33100  Udine, IT}

\begin{document}

\maketitle

\begin{abstract}
Early warning signals (EWS) based on critical slowing down (CSD) have been proposed as model-independent indicators of approaching ecological regime shifts and may help anticipate honeybee colony collapse. However, whether theoretically expected signals can be reliably observed and provide useful prediction in real-world systems remains uncertain. Here, we combine structural analysis, mechanistic modeling and longitudinal observations to assess the dynamical plausibility, detectability and predictive accuracy of CSD-based EWS for honeybee colony collapse.
Our results show that the presence of EWS is not equivalent to successful prediction of the transition. The analyzed models exhibit regime shifts accompanied by CSD, but detecting these signals in finite and noisy observations requires longer and more densely sampled time series than typically available for practical monitoring. Moreover, in empirical temperature records, CSD-like trends were detectable but provided limited discrimination between colonies that subsequently collapsed and those that survived. Declining mean temperature provided slightly greater discrimination than conventional CSD indicators, although its predictive value remained modest. Non-Gaussian correlated noise and the imperfect correspondence between hive temperature and colony strength further constrained EWS performance. These findings show that EWS can exist at the dynamical level without becoming useful predictors at the observational level, at least for our case study, thus cautioning against treating CSD-based indicators as generic predictors of ecological collapse.
\end{abstract}

\section{Introduction}
Honeybee (\textit{Apis mellifera}) colonies can experience abrupt declines in fitness and may collapse \cite{berenbaum2014bees, bacandritsos2010sudden}, with potentially relevant consequences for biodiversity, agriculture, and food security \cite{klein2007importance, pontarp2024evolutionary}; in particular, colony losses associated with infestations of the ectoparasitic mite \textit{Varroa destructor} and the \textit{Deformed wing} virus, which can make colonies collapse within one or two years \cite{nazzi2012synergistic}, are regarded as the biggest threat for the apicultural industry and the major cause of honeybee colony collapse \cite{genersch2010german}. Anticipating such transitions to collapse could therefore provide valuable opportunities for earlier intervention. The advent of affordable methods to assess features (such as temperature), that can be more or less related to the health status of a honeybee colony \cite{vsabic2025buzzing, colin2025temperature}, opened up new perspectives in this respect.
However, an actionable warning must appear sufficiently before collapse and be reliably measurable and interpretable; alerting for an impending collapse using observational data is thus particularly challenging as ecological systems are subject to multiple interacting stressors, environmental variability and seasonal changes.

\begin{figure}
    \centering
    \includegraphics[width=0.8\linewidth]{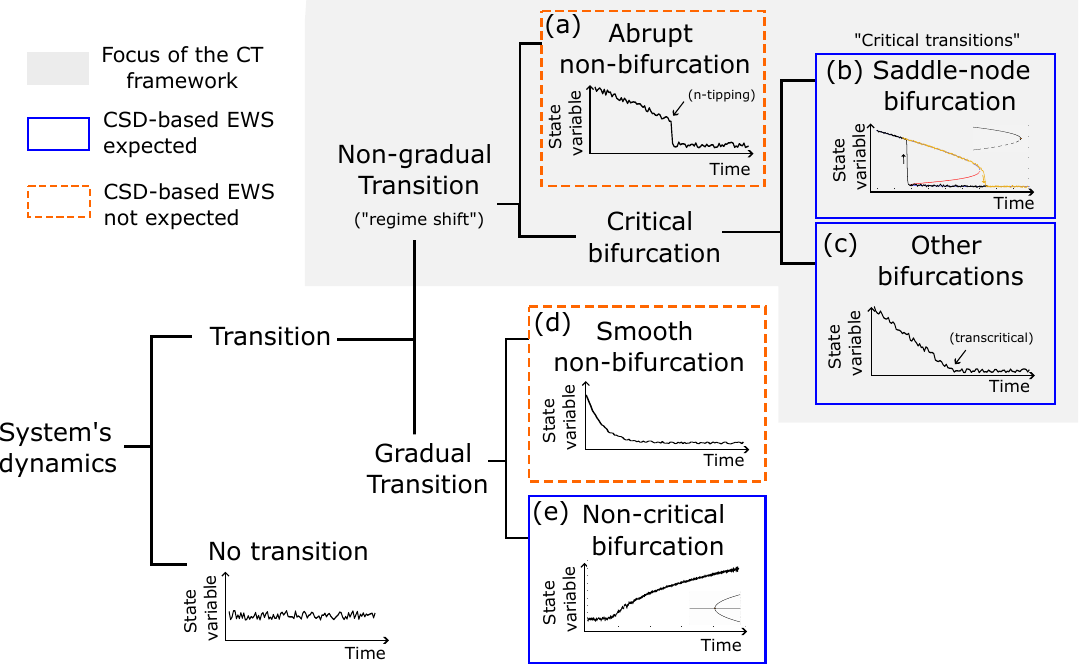}
    \caption{\small \textbf{Classification of transitions typically involved in dynamical EWS studies, with examples of time series.} Note that ``regime shifts'' broadly refer to abrupt changes between system states \cite{dakos2015resilience}, while ``critical transition'' may either refer to the general analysis framework (which encompasses the cases within the shaded area), or specifically to RS caused by saddle-node (SN) bifurcations, in panel (b) \cite{scheffer2012anticipating,proverbio2023systematic, o2023early}. The latter are of particular concern in ecology as they can yield catastrophic changes and hysteresis (see example) that complicate management and restoration \cite{Scheffer2009c,thompson2011predicting, bestelmeyer2011analysis}. Abrupt non-bifurcations (panel a) include noise-induced transitions (``n-tipping'' as in the example), rate-induced tipping (r-tipping), step-wise shocks, and more \cite{ashwin2012tipping, dakos2015resilience}. ``Other bifurcations'' (panel c) include the transcritical or subcritical cases \cite{kuehn2011mathematical}. All bifurcation-related transitions (b, c, e) imply the ``noisy b-tipping'' condition \cite{proverbio2023systematic}, i.e., noisy dynamics whose deterministic drift is driven by a bifurcation structure. CSD and EWS are linked with most bifurcation-associated regime shifts (blue frames) \cite{kuehn2011mathematical, proverbio2023systematic}, not only SN-driven. Instead, abrupt or smooth non-bifurcations are not theoretically associated with CSD-based EWS (dashed red frames). In all panels, time is shown on the x-axis as a proxy for a slowly varying bifurcation parameter $p$, typically assumed to evolve as $p = p_0 \pm a\cdot t$, with $a\ll 1$ \cite{berglund2006noise, kuehn2011mathematical}. This common assumption links bifurcation dynamics directly to the time series analysis used in later sections.}
    \label{fig:schemeCT}
\end{figure}

Early warning signals (EWS) based on Critical Slowing Down (CSD) have been proposed as a model-independent approach to anticipate abrupt regime shifts (RS; see Fig. \ref{fig:schemeCT} for an overview) in diverse natural systems \cite{scheffer2012anticipating}. CSD describes the progressive slowing of recovery from perturbations as a system approaches certain critical transitions through points of no return (``tipping points'') \cite{ashwin2012tipping,berglund2006noise,kuehn2011mathematical}. As recovery becomes slower, present dynamics become more correlated with its past and fluctuations persist for longer and accumulate \cite{kubo1966fluctuation}, leading to characteristic increases in lag-1 autocorrelation and variance \cite{kuehn2011mathematical, proverbio2022c}. Increases in other statistical indicators, including skewness, kurtosis, spectral properties and entropy, have also been proposed as EWS \cite{guttal2008changing,xie2019can,bury2020detecting,proverbio2023systematic}. Contrary to typical ecological analysis, that focuses on identifying system-specific stressors such as chemical, parasites and pathogens \cite{frizzera2023varroa, bekic2014honey, nazzi2012synergistic}, EWS analysis looks at systems' variables, parameters and noise \cite{booton2017stress} and seeks generic statistical patterns that signal proximity to the tipping point.  
These signals are attractive because they can potentially be extracted from observational time series without requiring a detailed mechanistic model \cite{scheffer2012anticipating}; they do not explain the underlying mechanisms, but aim to give decision-makers time to intervene and prevent undesirable outcomes \cite{lenton2011early, Alibakhshi2025}.

The applicability of such indicators to real natural systems, however, has shown mixed success \cite{o2023early,wagner2015false, wen2018one, gsell2016evaluating, proverbio2022performance} and is conditional to several factors. First, CSD is expected only for certain types of transitions \cite{proverbio2023systematic, dakos2015resilience}: those involving mildly perturbed dynamics and slow and gradual parameter changes near low-dimensional bifurcation points (\cite{kuehn2011mathematical}, see also Fig.~\ref{fig:schemeCT}). Sudden shocks, rapid parameter changes, or stochastic effects, such as multiplicative or correlated noise, may not yield CSD or display uninterpretable warning signals \cite{wang2012flickering, boettiger2013no, clements2016rate, qin2018early, kuehn2022warning}; nonlinearities and high system dimensionality can also make early detection more challenging \cite{romano2018analysis, weinans2021evaluating}. 
Second, the relevant state variable, or a sufficiently informative proxy, must be observable with adequate temporal resolution and precision. Finite and irregular sampling, measurement noise, seasonal forcing or temporally correlated environmental fluctuations can obscure or mimic the statistical signatures of CSD \cite{wagner2015false,morr2024anticipating,southall2020prospects,tredennick2022anticipating}. Warnings also must be detected within meaningful time frames: in ecology, EWS must emerge before predictable changes such as seasonal or cyclic event, rather than occurring at any point in the future \cite{dakos2008slowing}. 
Third, even when an indicator increases as theoretically expected, this does not necessarily imply useful prediction: similar changes may occur in systems that do not subsequently undergo a transition, generating false warnings \cite{Kefi2013a,zhang2015predictability,proverbio2024early}. Thus, the presence of a theoretically expected EWS and its practical predictive value are distinct questions.
These considerations motivate a distinction between three requirements for applying EWS to ecological monitoring. Dynamical plausibility asks whether the biological system can exhibit a transition for which EWS are theoretically expected; detectability asks whether the relevant dynamical signatures can be recovered from available measurements given their sampling frequency, noise structure, temporal extent and relationship to the underlying state; predictive utility asks whether observing such a signal provides sufficiently specific and timely information about a subsequent transition, beyond what is provided by ordinary changes in system state or environmental conditions. 
This distinction is important because an indicator can increase as theoretically expected near a transition without necessarily providing sufficiently specific or timely information that the system will undergo that transition \cite{Alibakhshi2025, boettiger2012early, spears2017ecological}.

These issues are particularly relevant to honeybee colonies. CSD-based EWS, estimated from frequently sampled time-series of hive temperature, have recently been proposed as a means of assessing colony condition and identifying those potentially at risk of collapse \cite{colin2025temperature}. However, it remains unclear whether the dynamical conditions required for EWS are present across relevant levels of honeybee organization, how the structure of the underlying transition affects the expected signals, and whether the available temperature measurements provide sufficient temporal and statistical information to detect them reliably and with sufficient lead time. In addition, honeybee colonies undergo seasonal changes, and internal temperature may reflect several interacting processes rather than a single measure of colony fitness. Consequently, increases in variance, in autocorrelation or in other typical CSD indicators extracted from hive temperature cannot automatically be interpreted as evidence of an approaching collapse and must be backed up by theoretical interpretation.

In this work, we systematically assess if and how CSD-based EWS are theoretically and practically actionable to anticipate colony dynamics. We first examine whether single-bee health can, in principle, exhibit multistability and regime shifts using structural analysis on a qualitative interaction network. We then analyze mechanistic models of colony dynamics using bifurcation theory to determine the structure of the transitions they can exhibit and to establish whether CSD-based EWS are theoretically expected. Next, using stochastic simulations, we investigate how sampling resolution, observation length and stochastic variability affect the detectability of these signals in synthetic data. Finally, we analyze longitudinal measurements of internal hive temperature and assess both the temporal behavior of conventional EWS and their ability to discriminate colonies that subsequently collapsed from those that survived.
We thus progress from establishing the dynamical conditions under which EWS can arise to assessing their practical applicability for real colonies. 

\section{Results}

\subsection{Pre-screening of critical transitions in single-bee health}
\label{sec:structural_analysis}

In natural systems where historical data are scarce and mechanistic models are unavailable, researchers often assume the presence of a regime shift and interpret trends in statistical indicators as early warning signals (EWS) \cite{scheffer2012anticipating, colin2025temperature}. However, this approach relies on the strong assumption that the system actually undergoes a regime shift, which critically affects how EWS are interpreted and applied and should therefore be carefully evaluated. Recent studies \cite{o2023early,ditlevsen2010tipping} have shown that several widely accepted regime shifts in ecology and climate science may not correspond to true critical transitions, a discovery that helps explain the poor performance of EWS in those cases. Assessing whether a system can theoretically undergo a regime shift is therefore a paramount first step to correctly interpret and apply EWS.
We therefore first ask whether the candidate biological variables have a dynamical structure compatible with regime-shift theory, before evaluating statistical indicators in observations. This provides a pre-screening step that distinguishes three situations: a transition may be dynamically possible and theoretically associated with EWS; a transition may be possible but the relevant state variable may be unobservable; or the candidate variable may not exhibit the type of transition for which conventional EWS are expected.

When quantitative data and detailed system knowledge are lacking, researchers have explored whether qualitative features, such as feedback structures or specific interaction networks, can indicate the potential for tipping points \cite{kefi2016can}. The problem of understanding whether a certain property (such as stability of a regime) is preserved or not, based on information about the structure of a system (i.e., its set of interactions and links) and regardless of quantitative details such as parameter values or functional forms, has a long tradition in control theory, which has established conditions under which systems are guaranteed to be stable, oscillatory, or multistable; see \cite{BG2021structural, PKG2025} and references therein. 

\begin{figure}
    \centering
    \includegraphics[width=0.9\linewidth]{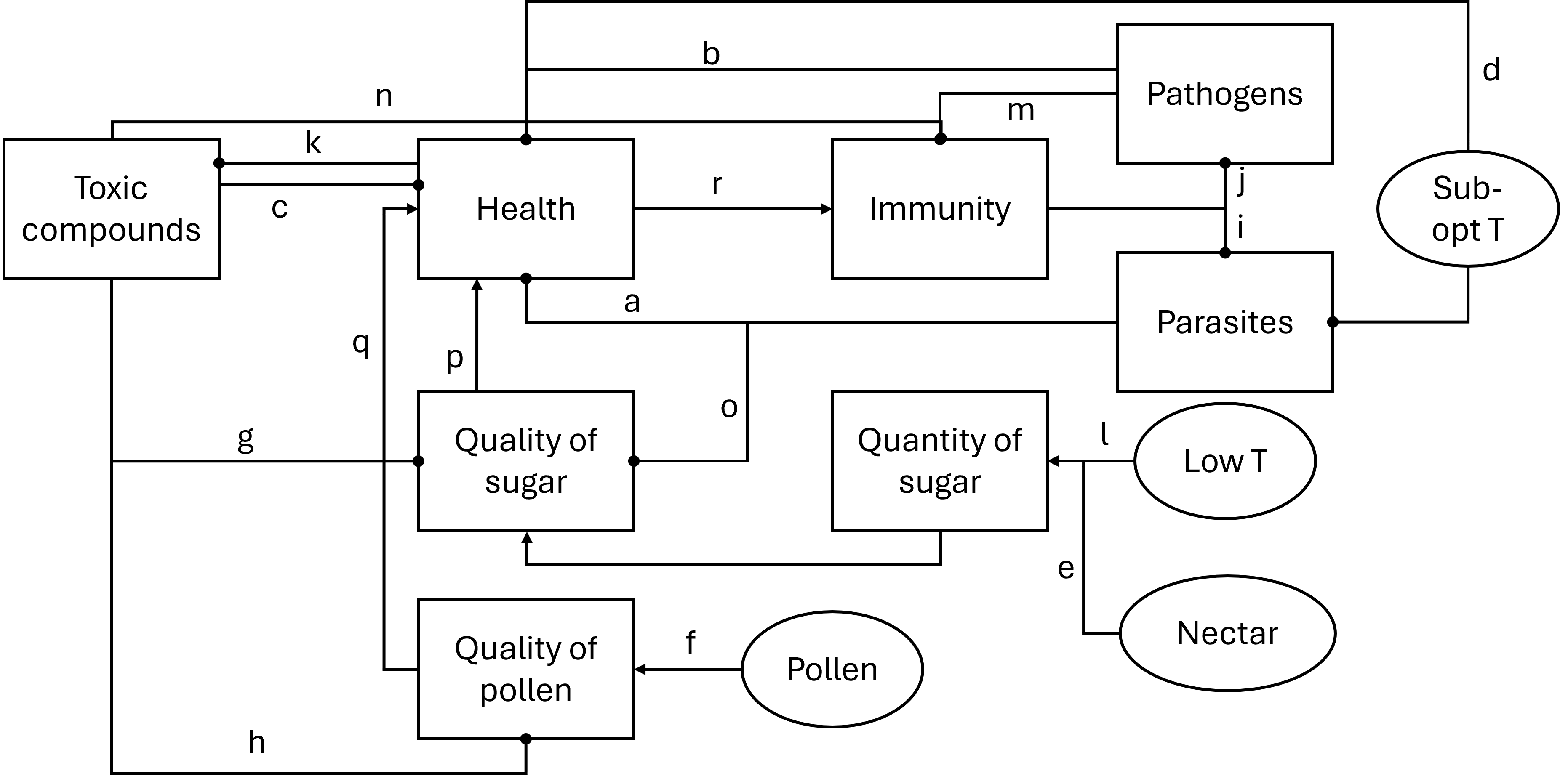}
    \caption{\small \textbf{Scheme of the interactions involved in a single-bee health model.} Pointed arrows indicate positive interactions (enhancement, +); bullet arrows indicate negative interactions (suppression, -). Negative self-influences are also assumed but not shown for health, toxic compounds, pathogens and parasites \cite{breda2022deeper}. Circled elements are environmental inputs. ``Sub-opt T'' stands for ``suboptimal temperature''; ``Low T'' for ``Low temperature''. Figure adapted from \cite{breda2022deeper}.}
    \label{fig:scheme_bee}
\end{figure}

Since alternative regimes must exist for a regime shift to occur, multistable systems are natural candidates to undergo RS. 
We first apply this approach to individual-bee health. 
As commonly occurs in ecology \cite{kefi2016can}, only graph-based or phenomenological models are available to describe the dynamical influences on single-bee health, making qualitative approaches particularly useful. Specifically, we consider an interaction network proposed in the literature \cite{breda2022deeper}, which describes the factors influencing bee health (such as temperature, nutrition, and pathogens), along with the sign (positive, as enhancer/promoter, or negative, as repressor/suppressor) of their effects; see Fig.~\ref{fig:scheme_bee}. Then, we apply two qualitative tests, also used in \cite{PKG2025} for other natural systems, to assess whether single-bee health dynamics could, in principle, be multistable and exhibit regime shifts.

The first test examines qualitative stability \cite{edelstein2005mathematical}. Each interaction is assigned a sign (+ for positive influences, - for negative influences, or 0 for no influence), and we assess whether the system is stable regardless of parameter values. If qualitative stability holds, regime shifts can be excluded (though the converse is not guaranteed). By constructing the sign matrix $A=\{a_{ij}\}$ of the interactions between element $i$ and $j$ and applying known criteria (see \cite[Sec. 6.5]{edelstein2005mathematical} and Supplementary Sec. S1), we find that the system violates conditions for qualitative stability due to the presence of positive feedback loops on the links \textit{c-k} and \textit{m-j} (see Fig.~\ref{fig:scheme_bee}). Therefore, the system is not guaranteed to be stable, and non-smooth or abrupt transitions cannot be ruled out.

The second test evaluates whether the system is a candidate for multistability. In monotone systems, multistability is associated with the presence of positive feedback cycles \cite{blanchini2014structural}: if all cycles are positive, the system is a strong candidate; if at least one is positive, it is a weak candidate. In the considered network (Fig.~\ref{fig:scheme_bee}), all non-self feedback cycles are positive (recall that $++ = +$, $--=+$, and $+-=-$), indicating that the system is a strong candidate for multistability. Furthermore, all loops and paths between stressors (Toxic Compounds, Pathogens, and Parasites) and Health are also positive, suggesting the possibility of bistability according to the criteria in \cite{angeli2004detection}. This implies that, for certain parameter ranges, the system may undergo saddle-node bifurcations and associated critical transitions.

Overall, these structure-based analyses show that single-bee health can, under the model assumptions, exhibit abrupt regime shifts. This establishes that regime-shift dynamics are theoretically possible at the individual level and provides a dynamical basis for considering EWS for this variable. Incorporating specific functional forms for interactions can further clarify the type of transitions involved. For example, previous work \cite{breda2022deeper} suggests that pathogen effects on bee health may follow Michaelis–Menten-type dynamics, which can lead to saddle-node bifurcations.
However, theoretical plausibility is not sufficient for practical monitoring. To constitute a useful EWS observable, the relevant state variable must also be measurable with sufficient temporal resolution and signal-to-noise ratio to resolve the recovery dynamics preceding the transition \cite{southall2020prospects, carpenter2011early}. Individual-bee health is not currently available at the temporal resolution required for this analysis. Actually, to date, any reliable measure of that variable implies the killing of the honey bee, making timely monitoring practically impossible. Future technologies may fill this gap and leverage our results to track individual health. As a result, although CSD-based EWS may be theoretically relevant, they are unlikely to be useful for real-world monitoring of individual bees. 

On the other hand, for most practical purposes, the biologically relevant scale is the colony, without which the individual or any extrapolated colonial organism cannot survive. Moreover, the important ecological service of pollination is a collective enterprise of honeybees, orchestrated by the waggle dance and carried out by the colony as a whole. In the next section, we therefore use colony-level temperature as an observable of its health and subsequently assess whether its dynamical structure and measurement properties permit the extraction of EWS.

\subsection{Model-based analysis of EWS for bee colonies}
\label{sec:models_trans}

Unlike the single-bee health model, colony-level dynamical models of honeybee fitness are more detailed and include explicit functional forms describing interactions among variables. Notably, such models often include mortality terms, due to pathogens, that can be explained by the bistability present at the individual insect level \cite{breda2022deeper}, which was analyzed in the previous paragraph. The phenomenological models are then amenable to direct analysis using tools from bifurcation theory, which can determine whether bifurcation-driven transitions can occur and identify their type.

To study colony dynamics and potential early warning signals (EWS), we first consider a minimal yet informative model from \cite{khoury2011quantitative}: 

\begin{align}
    \dot{H} &= L \frac{H+F}{w+H+F} - H \left( \alpha - \sigma \frac{F}{H+F} \right) \nonumber \\ 
    \dot{F} & = H \left( \alpha - \sigma \frac{F}{H+F} \right) - mF \, .
    \label{eq:model}
\end{align}

\begin{figure}
    \centering
    \includegraphics[width=0.9\linewidth]{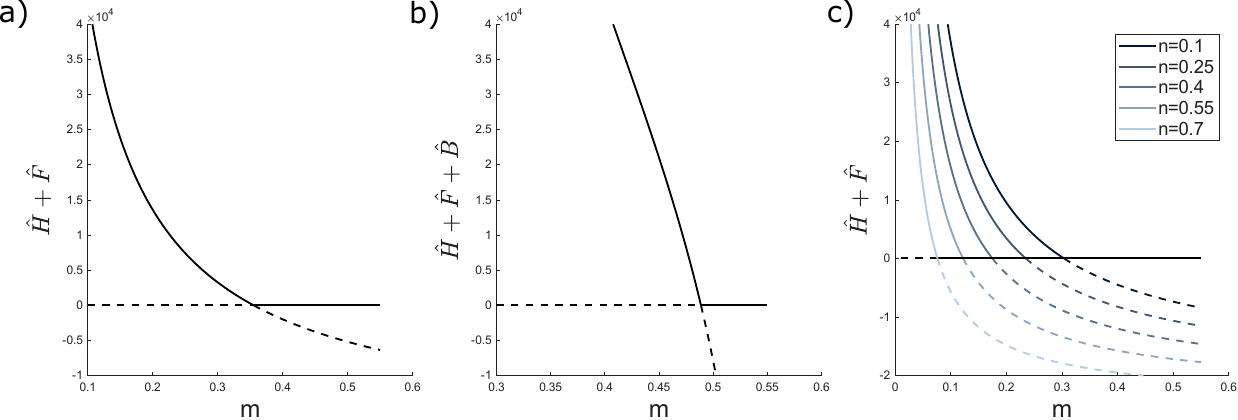}
    \caption{\small \textbf{Bifurcation diagrams for the colony fitness models:} a) the minimal model \eqref{eq:model}; b) the extended model \eqref{eq:model2} accounting for food foraging and consumption; c) the extended model \eqref{eq:model_breda} accounting for additional mortality mechanisms on young bees, captured by different values of the parameter $n$. The diagrams show colony strength against the bifurcation parameter $m$. Solid lines identify stable equilibria, dashed lines are for unstable equilibria. Details about the models, the calculation of their equilibria and of their critical value $m_c$ are in Methods.}
    \label{fig:bif_diags}
\end{figure}

Here, $H$ is the number of hive bees, $F$ is the number of foragers, $H+F$ is the total colony strength, $L, \sigma, \alpha, w$ are model parameters (respectively, the queen's egg laying rate, the reduction of $H$ recruitment if $F$ increases, the maximum rate at which $H$ turns into $F$, and the rate at which $L$ is reached as $H + F$ increases). The foragers' death rate $m$ is a bifurcation parameter: at a critical value $m_c$, the two equilibria that exist exchange stability (see details in Methods), indicating a transcritical bifurcation \cite{strogatz2018nonlinear}; see also Supplementary Fig. S1 for the phase plane of the model, for three different values of $m$ ($m<m_c$, $m>m_c$ and $m=m_c$). As $m$ increases from low values toward $m_c$, the stable nonzero equilibrium (representing a healthy colony) decreases until it collapses to zero for $m \geq m_c$; see the bifurcation diagram in Fig.~\ref{fig:bif_diags}a.
Similar behavior is observed in extended versions of model \eqref{eq:model}, that incorporate additional factors such as food dynamics \cite{khoury2013modelling} or increased mortality due to pathogens and environmental stressors \cite{breda2022deeper} (see Eq. \eqref{eq:model2} and \eqref{eq:model_breda} in Methods for details): as shown in Fig.~\ref{fig:bif_diags}b-c, the same transcritical structure is preserved also in these models. Apart from quantitative details, under the reasonable assumption that bee mortality changes slowly compared to typical population dynamics \cite{russell2013dynamic}, the transcritical bifurcation consistently describes the fundamental approach to colony collapse in this class of models. This can thus be interpreted as a regime shift, although not of the saddle-node type often assumed in the literature \cite{colin2025temperature}. Since transcritical bifurcations are still associated with CSD (recall Fig.~\ref{fig:schemeCT}), EWS based on CSD may, in principle, be observed, although they have different scaling properties compared to the saddle-node case \cite{proverbio2022performance}. \\

Having established that EWS may theoretically be measured from time-series data of honeybee colonies, the next step is to evaluate their practical detectability. To this end, we perform numerical simulations of the stochastic version of model \eqref{eq:model}, incorporating Gaussian white noise to reflect environmental variability \cite{kuehn2011mathematical, southall2020prospects, drake2026no} (see details in Methods). We employ numerical simulations to account for nonlinearities and investigate if EWS can be observed also far from the critical point, extending analytical results for transcritical-based dynamics \cite{proverbio2022performance, o2018stochasticity} that rely on a linearization close to the bifurcation point. For each value of the bifurcation parameter $m$, we simulate the system, compute statistical indicators (such as variance and lag-1 autocorrelation), and assess their trends using the Kendall’s $\tau$ coefficient; see details in Methods. Statistical significance is evaluated to determine whether these trends can serve as reliable EWS.

We focus on two key aspects relevant to real-world monitoring: the length of available time series ($L_D$) and the sampling frequency ($\rho$). Two scenarios are considered for $L_D$. The first is an idealized case with very long time series up to 10,000 days (over 27 years), allowing all indicators to reach statistical significance. While useful as a benchmark, this scenario is unrealistic for practical applications and neglects seasonal dynamics. The second scenario is more realistic, using one-year-long time series. For the sampling frequency $\rho$, we consider a wide range, from high-frequency measurements (every 0.1 days, comparable to temperature monitoring in hives, see the next section) to very sparse sampling (up to 200 days), including common intervals such as daily, weekly, and monthly observations.

\begin{figure}
    \centering
    \includegraphics[width=0.75\linewidth]{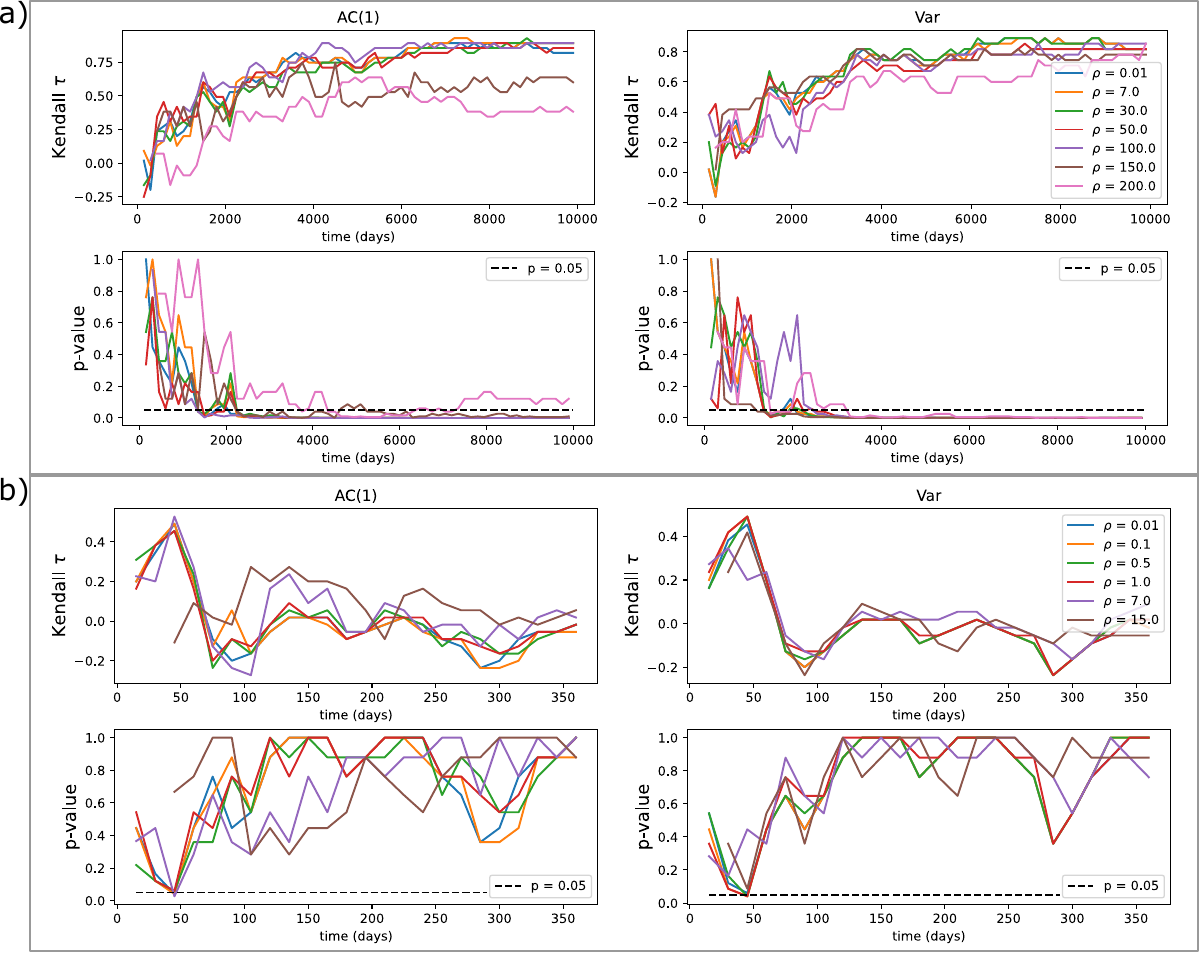}
    \caption{\small \textbf{EWS computed from simulations.} Kendall $\tau$ and its corresponding p-value are extracted from trends of AC(1) and Var, at different sampling resolutions $\rho$ (color-coded), considering (a) the full simulated time series with $L_D=10,000$ days, or (b) a shorter time series with $L_D=365$ days.}
    \label{fig:EWS-sim}
\end{figure}

Fig.~\ref{fig:EWS-sim} shows the $\tau$ values computed from trends of Var and AC(1) over time, along with their significance quantified with p-values compared with a significance level of $0.05$ (dashed line). Fig.~\ref{fig:EWS-sim}a is for the case $L_D=10,000$, Fig.~\ref{fig:EWS-sim}b for $L_D=365$, for various sampling resolutions $\rho$. We focus on EWS computed using Var and AC(1) since they are the most common indicators used in literature, and those least sensitive to model and sampling characteristics  \cite{proverbio2022performance}. Details about computation of the statistical indicators and of $\tau$ and the p-values are in Methods. An additional indicator, the Shannon Entropy proposed in \cite{proverbio2023systematic} is also tested, but it provides very little information, as reported in Supplementary Sec. S2. 
The results highlight clear limitations of EWS. For realistic time series of $L_D=365$, no statistically significant EWS is detected, regardless of $\rho$, except for brief and non-persistent fluctuations. In contrast, in the idealized long-term scenario, significant EWS based on variance and autocorrelation do emerge: when sampling is sufficiently frequent ($\rho < 100$), these signals become detectable after approximately 1,600 days (about four years). Although this provides some advanced warning relative to the full time series, it is still far longer than the typical lifespan and seasonal dynamics of honeybee colonies.

Overall, this analysis shows that, while colony-level dynamics are theoretically compatible with CSD-based EWS, practical constraints severely limit their usefulness, as the time required for indicators to become statistically significant is much longer than the relevant ecological time scales. As a result, even though EWS may exist in principle, they are unlikely to provide timely or actionable information for monitoring colony health in real-world conditions. Going one step further, the next section tests this \textit{in silico} result using empirical data.

\subsection{Empirical observability of EWS from temperature data}

The preceding analyses show that the conditions for early-warning signals are present at the level of the modeled dynamics: the colony models exhibit a critical transition, and conventional indicators such as variance and lag-1 autocorrelation can increase as the transition is approached. We therefore next ask whether these theoretically expected signals remain observable under realistic sampling and measurement conditions and, ultimately, whether their observation provides useful information about subsequent colony collapse. This distinction is important because the presence of an EWS is not equivalent to successful prediction of the transition.

Direct, high-frequency measurements of colony strength are currently unfeasible, making it hard to satisfy data quantity and quasi-steady assumptions for EWS analysis. In fact, current methods involve visual inspection of each of the hives' combs at night and by a trained observer (see Methods); the procedure is invasive and very time consuming \cite{guzman2025standard}. Continuous recording of the flight activity, colony weighting or other automatic methods are not yet sufficiently reliable, due to interferences of other factors such as meteorological conditions affecting foraging or the weight of the nectar stored in the hive. As an alternative, recent studies propose using internal hive temperature $(T_{\text{int}})$ as a proxy for colony health \cite{colin2025temperature}. In fact, bees actively regulate hive temperature near $35$ $^\circ$C in summer, while winter temperatures are lower and more variable, reflecting colony thermoregulation linked to brood development and colony functioning \cite{southwick1987temperature, minaud2024temperature}. Stable, elevated temperatures may therefore indicate higher fitness and stronger colonies, and lower temperatures may indicate decreasing fitness (rather than being a cause of colony failure). Importantly, $T_{\text{int}}$ can be measured non-invasively and remotely at high frequency, making it suitable for time series analysis.

Here, we analyze empirical $T_{\text{int}}$ data together with colony strength measurements, to test the practical applicability of EWS extracted from it. We use data from a controlled study conducted in 2018 in Northern Italy \cite{frizzera2023varroa}, involving two groups of five hives under similar environmental conditions (see details in Methods). One group was subjected to higher mortality due to \textit{V. destructor} infestation and eventually collapsed, while the other got treated and remained nearly uninfested throughout the experiment. Colony strength and infestation levels were sampled periodically but with low frequency, while internal temperature was recorded every 20 minutes thanks to sensors placed in the hives.
As detailed in Methods, we consider two datasets for $T_{\text{int}}$: $\mathcal{D}_1$ includes daily averaged temperatures from \cite{frizzera2023varroa}, while $\mathcal{D}_2$ preserves the original 20-minute resolution. External temperature $T_{ext}$ is also available from the local weather monitoring service. The datasets enable controlled comparisons between temperature dynamics, colony strength, and assessment of potential false or true warning signals, depending on sampling resolution.

\begin{figure}
    \centering
    \includegraphics[width=0.95\linewidth]{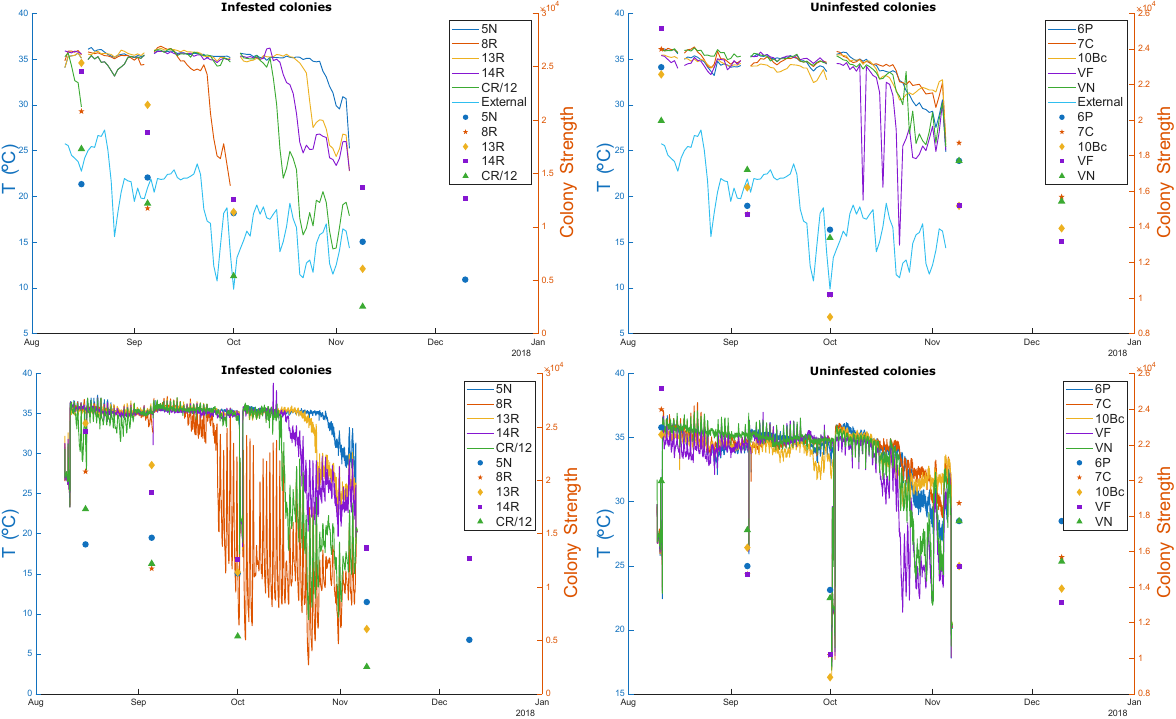}
    \caption{\small \textbf{Recorded time series of temperature (line plots) and colony strength (scatter plot; magnitude on the right y-axis).} Data are for infested (left column) and uninfested (right column) hives, from $\mathcal{D}_1$ (upper row) and $\mathcal{D}_2$ (lower row). For $\mathcal{D}_1$, the external recorded temperature $T_{\text{ext}}$ is also shown.}
    \label{fig:Temperatures}
\end{figure}

Fig.~\ref{fig:Temperatures} shows the raw temperature data (first row: $\mathcal{D}_1$; second row: $\mathcal{D}_2$) for the infested and uninfested hives (first and second column), together with the respective colony strength. 
Temperature patterns show strong seasonal structure. During summer and early autumn, $T_{\text{int}}$ is relatively stable around $34-35^\circ$C, then progressively declines toward lower values, either approaching collapse (see e.g. hive 8R) or in late autumn. Each hive exhibits a statistically detectable change point $t^*$, that marks a significant pattern shift (see Methods for details). For each hive, the corresponding $t^*$ are listed in Table~\ref{tab:changepoints} and visualized in Supplementary Figs. S3 (for infested) and S4 (for uninfested). However, such change points appear in both collapsing and healthy colonies, generally around mid-October, suggesting they are not unique indicators of fragility.

\begin{table}[h]
\centering
\small
\begin{tabular}{l|lllll|lllll|}
\cline{2-11}
                             & \multicolumn{5}{c|}{Infested}                                                                    & \multicolumn{5}{c|}{Uninfested}                                                                  \\ \hline
\multicolumn{1}{|l|}{Hive} & \multicolumn{1}{l|}{5N} & \multicolumn{1}{l|}{8R} & \multicolumn{1}{l|}{13R} & \multicolumn{1}{l|}{14R} & CR/12  & \multicolumn{1}{l|}{6P} & \multicolumn{1}{l|}{7C} & \multicolumn{1}{l|}{10Bc} & \multicolumn{1}{l|}{VF} & VN \\ \hline
\multicolumn{1}{|l|}{$t^*$}  & \multicolumn{1}{l|}{22/10} & \multicolumn{1}{l|}{18/09} & \multicolumn{1}{l|}{16/10} & \multicolumn{1}{l|}{12/10} & 07/10 & \multicolumn{1}{l|}{16/10} & \multicolumn{1}{l|}{14/10} & \multicolumn{1}{l|}{12/10} & \multicolumn{1}{l|}{09/10} & 17/10 \\ \hline
\end{tabular}
\caption{\textbf{Change points $t^*$ for each hive (dd/mm).} The year is 2018.}
\label{tab:changepoints}
\end{table}

The existence of change points and the decreasing patterns from Fig.~\ref{fig:Temperatures} are consistent with what was observed in previous works \cite{arias2026assessing}; however, they do not map straightforwardly onto colony strength. When looking at $\mathcal{D}_1$ ($\mathcal{D}_2$ presents notable daily fluctuations that need further pre-processing, as described in Methods), internal temperature and colony strength follow different trends: $T_{\text{int}}$ may remain stable while colony strength declines (e.g., hives 13R, 14R, 10Bc, VF between August and October), or decrease while colony strength remains stable or even increases (hives 14R, VF, VN between October and November). This suggests a nontrivial relationship between temperature and fitness, potentially shaped by colony strength, parasite effects on thermoregulation \cite{frizzera2023varroa}, and environmental influences. 

To test whether external temperature drives internal temperature changes, we apply Granger causality analysis (see Methods) using $\mathcal{D}_1$, which has the same sampling rate as $T_{ext}$. In nearly all hives, except for hive CR/12 (infested) and for hive VN (uninfested) for $t>t^*$, the results support the null hypothesis that $T_{\text{ext}}$ does not Granger-cause $T_{\text{int}}$; see also Supplementary Sec. S3.2 for details and fit results. Under the tested Granger-causality framework, we thus found no evidence that ambient temperature improves prediction of internal temperature; the latter may depend on additional factors linked to seasonality, precipitations, humidity, or availability of nutrients, bees physiology, or more. Disentangling all these effects require future dedicated experiments and models.

\begin{figure}
    \centering
    \includegraphics[width=0.95\linewidth]{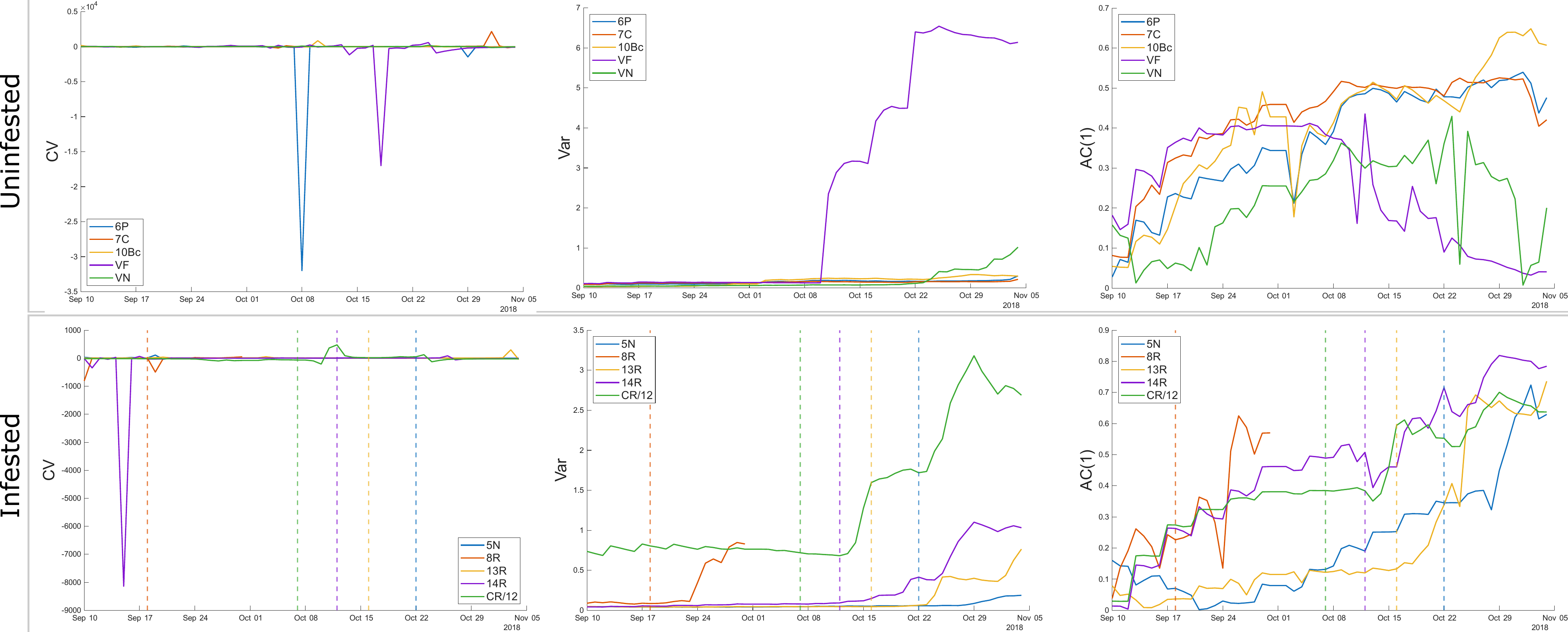}
    \caption{\small \textbf{Evolution of the considered summary indicators (CV, Var, AC(1)) for uninfested (top) and infested (bottom) hives from $\mathcal{D}_1$.} The dashed lines correspond to the identified change points $t^*$ for the infested hives, see Tab.~\ref{tab:changepoints} for values.}
    \label{fig:indicators}
\end{figure}

The analysis above also suggests that detrending $T^{\text{int}}$ by $T^{\text{ext}}$ is not informative; we thus employ a non-parametric LOESS approach to pre-process and detrend the time series before EWS calculation (see Methods for details). After pre-processing, we extract standard early warning indicators, i.e., coefficient of variation CV, variance Var, and lag-1 autocorrelation AC(1), using expanding windows in an online, non-retrospective fashion (see details in Methods). Results for the daily dataset $\mathcal{D}_1$ (Fig.~\ref{fig:indicators}) show that CV is largely erratic: aside from occasional drops, it remains mostly flat and is therefore excluded from further analysis. In contrast, variance and AC(1) often increase over time, particularly in infested hives, qualitatively consistent with predictions from the Critical Transitions framework. The same indicators computed on $\mathcal{D}_2$ are reported in Supplementary Sec. S4. 
Retrospectively, knowing the full evolution of the colonies and having access to the complete time series, these increasing trends could appear to signal colonies approaching regime shifts. However, practical monitoring has no access to future data, and requires that trends are statistically significant in real time and robust against fluctuations, such as those visible in Fig.~\ref{fig:indicators}. Fluctuating indicators alone do not guarantee reliable early warning to distinguish collapsing from non-collapsing colonies, and do not allow to avoid false alarms.

\subsection{Predictive utility of empirical EWS}

\begin{figure}
    \centering
    \includegraphics[width=0.7\linewidth]{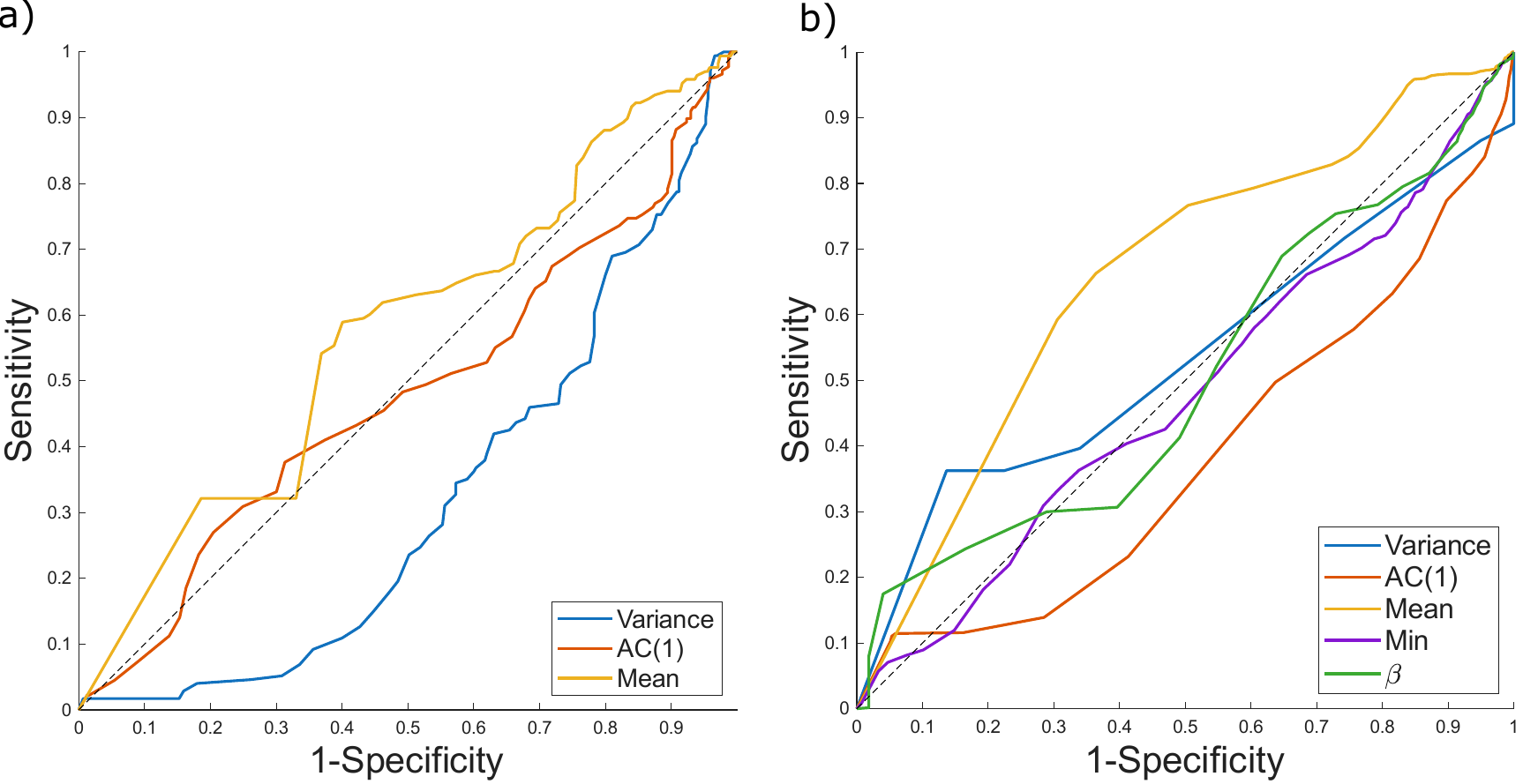}
    \caption{\small \textbf{ROC curves for the various indicators} computed from (a) $\mathcal{D}_1$ and (b) $\mathcal{D}_2$. The dashed diagonal line corresponds to a random guess classifier giving false and true positives with equal probability.}
    \label{fig:ROC}
\end{figure}

To evaluate performance quantitatively, we measure the consistency of increasing trends using the Kendall’s $\tau$ coefficient \cite{southall2020prospects, boettiger2012early} for each indicator and over the same expanding window (see Methods). We assess predictive skill, in terms of sensitivity and specificity of the signals, through Receiver Operating Characteristic (ROC) curves and Area Under the Curve (AUC) scores (see details in Methods). ROC curves are shown in Fig.~\ref{fig:ROC}, computed from $\mathcal{D}_1$ (panel a) and from $\mathcal{D}_2$ (panel b). In addition to Var, CV and AC(1), we also compute trends for the Mean, which should consistently decrease as the system slowly approaches its critical value. For the high-frequency dataset $\mathcal{D}_2$, which contains more information, we also examine additional indicators: daily minimum temperature (Min) and spectral slope $\beta$, see Methods for details.

Fig.~\ref{fig:ROC} suggests overall weak EWS performance. Most indicators perform similarly to, or worse than, a random classifier, implying that trends observed in collapsing colonies are often also present in healthy ones. They do not seem to be specific hallmarks of more fragile colonies, a fact that limits their specificity as warning signals. For $\mathcal{D}_1$, AUC values are 0.32 for Var, 0.48 for AC(1), and 0.57 for Mean: only the mean temperature is better than random classification. For $\mathcal{D}_2$, AUC values are 0.53 for Var, 0.38 for AC(1), 0.66 for Mean, 0.48 for Min, and 0.50 for the spectral slope $\beta$. Also in this case, only the Mean shows some predictive value. Interestingly, the Mean performs better in the high-frequency dataset than in the daily averaged one, likely because high-frequency sampling preserved information that was attenuated by daily averaging; whether this improves generalizable EWS performance requires validation in larger datasets.
Instead, the Min indicator contrasts with prior studies \cite{colin2025temperature}, performing poorly here (AUC = 0.48). Differences in environmental conditions, stressors, or measurement protocols may explain the discrepancy. More broadly, this inconsistency suggests that Min is not a robust transferable indicator without careful control of the context.

Taken together, these empirical results provide some support for bifurcation-related mechanisms underlying honeybee colony collapse, since warning signals are not entirely absent. However, there is generally weak discrimination between colonies that subsequently collapsed under the experimental conditions, and those that survived. The mean temperature showed the highest apparent discrimination (AUC = 0.66), although the small number of colonies suggests caution about this estimate and does not support reliable conclusions for routine application.
Thus, while empirical EWS partly support critical transition theory applied to honeybee colony dynamics, standard indicators here perform poorly for forecasting colony collapse in practice. The results therefore do not indicate an absence of dynamical signatures; rather, they demonstrate that their presence alone is insufficient to provide reliable prediction of colony outcome under the conditions examined here. Improved indicators, better mechanistic understanding, or richer data may be needed before EWS can become operational tools for monitoring colony fitness. 

\subsection{Interpretation and lead-time}

The limited performance of the tested early warning indicators can be understood from several perspectives. First, colony dynamics appears linked to a transcritical bifurcation, not a saddle-node one (\textit{cf.} Sec.~\ref{sec:models_trans}). Although transcritical systems can exhibit critical slowing down \cite{kuehn2011mathematical}, their mean often serves as a more direct predictor because it declines monotonically as the critical point is approached (see example in Fig.~\ref{fig:schemeCT}). Consistently, mean temperature was the best-performing indicator in both datasets and the only one reliably performing slightly better than random classification. Yet this comes with false positives, since similar mean declines can occur in colonies that never collapse.
Second, the results from Sec.~\ref{sec:models_trans} suggest that, to detect statistically significant EWS, this system may require very abundant, high-frequency data and very long time series, potentially longer than typical seasonal cycles. Third, the assumption of slowly varying control parameters (in this case, $m$) may be unrealistic. Colony mortality may change rapidly due to food shortages, weather anomalies, or worsening \textit{V. destructor} infestation, violating the quasi-steady assumptions underlying standard EWS methods. Future studies may deeply investigate this aspect, should more abundant data become available.

A further factor concerns noise. EWS often perform best for autoregressive systems driven by Gaussian white noise, and noise properties can strongly affect their reliability \cite{ proverbio2022performance,kuehn2022warning,o2018stochasticity}. Understanding which noise type is present in the system helps interpret the performance of the computed EWS. Using the high-frequency dataset $\mathcal{D}_2$, detrended fluctuations are analyzed before and after the detected change points listed in Table~\ref{tab:changepoints} (see details in Methods). The full results are reported in Supplementary Section S5: before the change point, where early warnings are most needed, the detrended fluctuations show noticeable skewness and sharp peaks around zero, which are inconsistent with Gaussian noise. After the change point, deviations from Gaussianity remain. 
Power spectral analysis further suggests that the system is dominated not by white noise but by red noise, with spectral slopes (explained in Methods) $\beta \sim 2$ across hives over the whole time series; details are in Supplementary Sec. S5. As non-Markovian, red and state-dependent noise can obscure EWS even in systems characterized by transcritical bifurcations and significantly reduce their applicability \cite{proverbio2023systematic, kuehn2022warning, o2018stochasticity}, this evidence contributes in explaining their weak performance. It also clarifies why spectral reddening, i.e., Power Spectral Density becoming closer to that of red noise as tipping points are approached \cite{bury2020detecting}, failed as a useful indicator (see $\beta$ in Fig.~\ref{fig:ROC} and Supplementary Sec. S4): the system already appears strongly red-noise dominated.

Despite these limitations, the ROC analysis allows to get an idea about $\tau_c^*$, the values of $\tau$ for which an ROC curve is the furthest up from the diagonal line, marking the current best tradeoff between sensitivity and specificity. One can then use persistent exceedance of these thresholds as criterion to raise an alert. Because the dataset contains a limited set of colonies, the ROC analysis results are mainly exploratory and not intended to provide an externally validated predictive threshold: $\tau_c^*$ is intended to better interpret the reliability of EWS in these data and provide a preliminary lead-time analysis, without being necessarily transferable to other contexts. For the best indicator, the Mean, $\tau_c^* = 0.66$ for $\mathcal{D}_1$ and $\tau_c^* = 0.76$ for $\mathcal{D}_2$. The first date $t_{\tau_c^*}$ after which $\tau>\tau_c^*$ persistently (at least 5 days in a row, to provide a more reliable signal) is reported in Table~\ref{tab:tau_best}, for each considered hive. Comparing these dates with those related to the change point in the raw data (Table~\ref{tab:changepoints}) supports the idea that some signals preceded detected temperature change points by several weeks; the small sample and uncertain collapse timing call for future studies about reliable estimates of predictive lead time.
In healthy colonies, warnings tend to occur closer to the change point, suggesting some ability to differentiate them from fragile hives.
The threshold computed from daily data $\mathcal{D}_1$ is relatively conservative and misses warnings in two hives, whereas this limitation disappeared with high-frequency data, again underscoring the value of dense sampling. Because the exact timing of colony extinction is unavailable and sample sizes are small, stronger conclusions about lead times, i.e., how much advance a warning provides with respect to the date of collapse, are not yet possible.

Overall, these results suggest that weak EWS performance is not a methodological limitation but may reflect key properties of the system: transcritical rather than saddle-node dynamics, rapid forcing, limited data, and strongly correlated non-Gaussian noise all work against standard CSD-based indicators.
Still, the possibility of generating warnings some weeks in advance, even with modest accuracy, may have practical value. For beekeeping, an imperfect alert raised a month early could still be useful if it enables preventive action. This points toward a pragmatic tradeoff between precision and timeliness; it also motivates future work on alternative indicators, richer datasets, and closer collaboration with practitioners to develop more operational warning systems.

\begin{table}[h]
\centering
\small
\begin{tabular}{l|lllll|lllll|}
\cline{2-11}
                             & \multicolumn{5}{c|}{Infested}                                                                    & \multicolumn{5}{c|}{Uninfested}                                                                  \\ \hline
\multicolumn{1}{|l|}{Hive} & \multicolumn{1}{l|}{5N} & \multicolumn{1}{l|}{8R} & \multicolumn{1}{l|}{13R} & \multicolumn{1}{l|}{14R} & CR/12  & \multicolumn{1}{l|}{6P} & \multicolumn{1}{l|}{7C} & \multicolumn{1}{l|}{10Bc} & \multicolumn{1}{l|}{VF} & VN \\ \hline
\multicolumn{1}{|l|}{$t_{\tau_c^*}(\mathcal{D}_1)$}  & \multicolumn{1}{l|}{NaN} & \multicolumn{1}{l|}{NaN} & \multicolumn{1}{l|}{29/09} & \multicolumn{1}{l|}{23/09} & 31/10 & \multicolumn{1}{l|}{11/09} & \multicolumn{1}{l|}{21/10} & \multicolumn{1}{l|}{01/11} & \multicolumn{1}{l|}{06/10} & 24/10 \\ \hline
\multicolumn{1}{|l|}{$t_{\tau_c^*}(\mathcal{D}_2)$}  & \multicolumn{1}{l|}{14/09} & \multicolumn{1}{l|}{24/08} & \multicolumn{1}{l|}{23/09} & \multicolumn{1}{l|}{23/10} & 20/09 & \multicolumn{1}{l|}{15/09} & \multicolumn{1}{l|}{23/10} & \multicolumn{1}{l|}{23/09} & \multicolumn{1}{l|}{23/10} & 20/09 \\ \hline
\end{tabular}
\caption{\textbf{The first date $t_{\tau_c^*}$ (dd/mm) after which $\tau>\tau_c^*$ persistently} (at least 5 days in a row), for each hive and for $\mathcal{D}_1$ and $\mathcal{D}_2$. The year is 2018. NaN means that no date was found: the warning is never raised for the given $\tau_c^*$. For $\mathcal{D}_2$, we show the day without the detailed hour, for comparison.}
\label{tab:tau_best}
\end{table}

\section{Discussion}

Generic early warning signals based on critical slowing down may help ecologists in their monitoring and management efforts, but their practical value remains debated. Motivated by the need to anticipate losses of honeybee fitness, we combined model-based and empirical approaches to determine at which level of biological organization a measurable regime shift exists and to test whether EWS can predict colony collapse. Using robustness analysis, we showed that honeybee systems can undergo regime shifts at both the individual and colony levels, supporting an assumption that is often made in literature but rarely tested. However, the existence of critical dynamics does not, by itself, imply that EWS will be detectable or predictive. Notably, our theoretical conclusions are generic and hold irrespectively of sample sizes. On the other hand, the required precision and frequency of assessments prevented us from performing the empirical EWS sensitivity analysis on a larger sample size; future studies may expand the sample and possibly refine the quantitative AUC values.

Focusing on honeybee colonies, we examined the theoretical and empirical performance of EWS. 
Three mechanistic models consistently indicated that colony loss can arise through a transcritical bifurcation. This provides a plausible alternative to the saddle-node structure that underlies much of EWS theory and highlights that SN regime shifts are not a universal feature of ecological systems. Because transcritical bifurcations can produce weaker or less immediate signals, identifying the underlying dynamical structure is important when assessing if EWS are actionable. In simulations, conventional indicators could emerge retrospectively, but their statistical significance was limited when evaluated online over realistic ecological timescales. Signals that appear only shortly before collapse, or are too weak relative to background variability, have limited operational value. Future works, considering more complex models, different noise types or parameter change rates, may further quantify the horizon needed to extract EWS and their significance in alternative settings. Overall, these results highlight that identifying the correct dynamical structure is essential before applying or interpreting EWS.
The empirical analysis reinforced this distinction between dynamical signals and useful prediction. Using internal hive temperature from collapsing and healthy colonies, we evaluated standard univariate indicators at two sampling frequencies, assessed their discriminatory performance using ROC analysis, and related warnings to change points in temperature dynamics. Internal temperature provided too imprecise a proxy for colony strength to support reliable EWS: most indicators contained little predictive information, with the main exception of the intuitive decline in mean temperature, which nonetheless only limitedly performs better than a random classifier and does not bring adequate precision for most practical applications. In this system, simple state-based measurements thus outperform theoretically motivated CSD indicators. Where feasible, we suggest future studies to measure colony strength directly and at higher temporal resolution rather than rely on temperature as a proxy. Direct estimates of mortality rates $m$, or other control parameters, would additionally allow to assess more directly the existence and magnitude of critical timescales. Experiments independently varying infestation intensity, colony strength and subsequent survival would further help determine whether temperature-derived indicators contain predictive information beyond direct measures of colony condition.

More broadly, our results emphasize the need to establish whether EWS are both theoretically justified and practically observable before applying them to real-world systems. A crucial distinction is between dynamical and predictive evidence. Variance and autocorrelation increased in both simulations and empirical records, consistent with processes capable of generating CSD. However, these changes were insufficiently specific to discriminate reliably between collapsing and surviving colonies. An indicator can therefore behave as expected near a transition without substantially increasing the probability that a transition is imminent, or without providing sufficient warning time for intervention. For ecological monitoring, specificity, timeliness and discrimination, not just the presence of a theoretically expected signal, ultimately determine predictive utility.
This distinction connects naturally to a Bayesian decision perspective on EWS \cite{Alibakhshi2025, rietkerk2025ambiguity}, in which the probability that a regime shift can occur must be considered before interpreting an observed warning signal (see Supplementary Sec. S6). This perspective motivates the analytical sequence adopted in this study. We first assess whether regime-shift dynamics are plausible in honeybee systems. The robustness framework applied here provides a qualitative first step toward this assessment and can be extended to other ecological systems. It presents an alternative to studies that use empirical observations to test whether a system had been subject to critical transitions in the past \cite{o2023early}, or to model-based analysis of tipping elements \cite{rietkerk2025ambiguity}, performed when mechanistic models exist. With sufficiently validated models and abundant data, probabilistic robustness approaches \cite{sutulovic2026gpc} could further quantify the probability of critical transitions rather than treating their existence as a binary property.
We then test whether the modeled transition generates CSD-based signals, examine if these signals can be detected under realistic sampling and noise conditions and, finally, whether their observation discriminates colonies that subsequently collapse from those that survive. Theoretical expectation, empirical detectability and predictive utility represent distinct levels of EWS assessment.

Despite the current limitations observed in EWS, our results provide several constructive directions. First, both modelling and empirical analyses indicate that higher sampling frequency can improve EWS performance by preserving informative temporal dynamics. Second, the predominance of transcritical dynamics in our colony models motivates systematic investigation of whether conventional EWS are generally weaker or less timely near transcritical than near saddle-node bifurcations \cite{proverbio2022performance,drake2026no}. Such systems may require dedicated warning indicators or alternative predictive approaches.
Also, although temperature appears here unsuitable for reliable anticipation of colony collapse, it may nevertheless support decentralized and low-cost monitoring of colony condition \cite{arias2026assessing}. Qualitative temperature dynamics can distinguish broad colony states: relatively stationary internal temperatures are characteristic of apparently healthy colonies (Fig.~\ref{fig:Temperatures}, approximately until October), whereas declining colonies show increasingly variable and decreasing temperatures, and collapsed colonies approach external temperature. This constitutes detection rather than prediction; nonetheless, detection can still be valuable for remote monitoring when direct colony inspections are difficult. For this purpose, simple direct measurements may prove more robust than complex summary statistics. Further work should clarify the relationship between colony strength and temperature, including its directionality, causal structure and functional form.

We focused on univariate indicators because they are directly interpretable and closely connected to dynamical theory. Multivariate indicators \cite{proverbio2023systematic,o2023early} may improve discrimination, although they generally augment rather than resolve weak univariate signals and introduce additional assumptions. Their evaluation is therefore an important direction for future work. Machine-learning (ML) approaches \cite{bury2021deep} may provide complementary predictive tools, but were beyond the scope of this study, which aimed to assess EWS specifically while retaining a mechanistic interpretation of their applicability. Moreover, many ML approaches rely on surrogate systems constructed around standard assumptions about critical slowing, noise and dynamics, and require system-specific finetuning \cite{bury2021deep}. These assumptions are unlikely to hold for hive temperature, which exhibits correlated and non-Gaussian variability; richer data would therefore be needed to determine whether ML can overcome the limitations identified here. Trait-based indicators \cite{clements2016including} and model-driven approaches may offer additional routes toward earlier prediction of colony collapse.

Overall, our findings show that EWS cannot be transferred uncritically across ecological systems. Their success depends on several conditions: regime shifts must be dynamically possible, the underlying bifurcation structure must generate sufficiently strong and timely signals, assumptions about timescales and noise must be appropriate, and the observed variable must adequately represent the relevant system state. To address these challenges, we provided a framework for qualitatively assessing whether regime shifts are plausible in a natural system and demonstrate its application to honeybee health and colony dynamics. We also showed, both theoretically and empirically, that EWS can exist at the dynamical level without becoming useful predictors at the observational level. Hence, interpreting EWS requires moving beyond identifying statistical changes toward assessing whether those changes are sufficiently specific, detectable and timely, and whether they meaningfully increase the probability of an impending transition.
EWS should therefore be viewed not as universal predictors, but as system-dependent tools whose usefulness must be demonstrated rather than assumed. Better mechanistic models, higher-resolution measurements and complementary predictive approaches may improve the early identification of vulnerable colonies and, more broadly, help bridge the gap between dynamical theory and actionable prediction in ecology.

\section{Methods}

\subsection{Stability and critical values of colony models}

The equilibria of the colony model \eqref{eq:model} are easily obtained \cite{khoury2011quantitative} by setting $\dot{H}=0, \dot{F}=0$, yielding 
\begin{equation}
    \hat{H}=\hat{F}/J \quad \quad \hat{F}=L/m - wJ/(J+1)
    \label{eq:equilibria1}
\end{equation}
with $J=\left(\alpha-\sigma-m+\sqrt{(\alpha-\sigma-m)^2+4\alpha m)} \right)/2m$. To find the critical parameter value $m_{c,1}$, we ask that $\hat{F}+\hat{H}>0$ (condition of non-zero colony strength) and solve for the bifurcation parameter $m$, obtaining 
\begin{equation}
    m_{c,1} = \frac{L}{2w\left(\alpha-\frac{L}{w}\right)}\left(\alpha+\sigma+\sqrt{(\alpha-\sigma)^2+4\sigma\frac{L}{w}} \right) \, 
    \label{eq:mc_khoury}
\end{equation}
which is ecologically meaningful if $\alpha > L/w$. The non-zero equilibria are globally asymptotically stable if $m<m_{c,1}$ \cite{khoury2011quantitative}. The critical value $m_{c,1}$ is a transcritical point (assessed symbolically), after which only the equilibrium $(0,0)$ exists and is stable. \\

More complex models of honeybee colony dynamics show qualitatively similar structures. The model in \cite{khoury2013modelling} extends Eq. \eqref{eq:model} by adding harvesting and consumption of food into the dynamical equations. Note that, since the interaction between foragers and food is not monotonous (see Fig. 2 in \cite{khoury2013modelling}), the approaches used in Sec.~\ref{sec:structural_analysis} cannot be applied here, and a complete dynamical analysis must be conducted. The complete model is:
\begin{align}
    \dot{B} & = L \frac{H}{H+v} \cdot \frac{f^2}{f^2+b^2} - \phi B \nonumber \\
    \dot{H} &= \phi B - H \left(\alpha_1 - \alpha_2\frac{b^2}{f^2+b^2} - \sigma \frac{F}{H+F} \right)  \nonumber \\
    \dot{F} &=   H \left(\alpha_1 - \alpha_2\frac{b^2}{f^2+b^2} - \sigma \frac{F}{H+F} \right) - mF \nonumber \\
    \dot{f} &= cF-\gamma_A(F+H)-\gamma_B B \, ,
    \label{eq:model2}
\end{align}
where the additional variables are brood $B$ and food stores $f$. The additional parameters are $v$, which determines the rate at which the food-induced capping $S=H/(H+v)$ is reached, $\phi$ (the pupation rate of brood), $\alpha_1 = \alpha$, $\alpha_2$ that governs the transition of $H$ into $F$ when food stores are low, $\gamma_A$ and $\gamma_B$ (in [g/day], the food stores consumed by Adult $F$ and $H$ bees, or by Brood), and $c$, the average amount of food collected by each forager per day. The constant $b$ shapes the food-dependent terms as a function of food stores. The model in \cite{khoury2013modelling} considers a delayed term $B(t-\bar{\tau})$, but the authors observe the same bifurcation structure regardless of $\bar{\tau}$. 
When $\bar{\tau}=0$, the model has an asymptotically stable non-zero equilibrium given by \cite{khoury2013modelling}: 
\begin{align}
    \hat{f} &= b \sqrt{\frac{\alpha_2}{\frac{\sigma}{P+1}+\frac{m}{P}-\alpha_1}-1} \, \quad  \quad &\hat{H} = P\hat{F} \, \nonumber \\
    \hat{F} &=  \frac{L\hat{f}^2}{m(\hat{f}^2+b^2)} \, \quad  \quad &\hat{B} = \frac{m\hat{F}}{\phi} \, ,
    \label{eq: equilibria2}
\end{align}
where 
\begin{equation*}
    P = \frac{c}{\gamma_A}-1-\frac{\gamma_B m}{\gamma_A \phi} \, 
\end{equation*}
and under the conditions that $\alpha_1<\sigma/(P+1)+m/P$ and $\alpha_2>\sigma/(P+1)+m/P-\alpha_1$, if $m<m_{c,2}$ (see below).
The functional forms for $F$ and $H$ of the original model \eqref{eq:model} can be recovered if food is assumed to increase indefinitely, $f\to \infty$ for $t\to\infty$.
Also in this case, the bifurcation diagram $(m; \hat{F}+\hat{H}+\hat{B})$ follows a transcritical structure (see Fig.~\ref{fig:bif_diags}b) and there exists a critical value $m_{c,2}$ such that the colony is alive if $m<m_{c,2}$ and goes extinct if $m\geq m_{c,2}$. To get $m_{c,2}$, we ask that $\hat{H}>0$, so that
\begin{equation}
    m_{c,2} = \frac{\phi}{\gamma_B}\left(c-\gamma_A \right) \ .
\end{equation}
In this case, $m_{c,2}$ does not depend on bee demographic parameters ($L$, $w$, $\alpha$, $\sigma$), but on the foraging parameters and on the pupation rate $\phi$. Despite the quantitative details and the ecological meaning, the dynamical structure characterized by a transcritical bifurcation is nonetheless maintained for a given $c$: this observation supports the use of EWS that are generic for a bifurcation type instead of being dependent on functional details. \\

Finally, the model from \cite{breda2022deeper} extends \eqref{eq:model} by including an additional death rate of young bees due to, e.g., environmental conditions, \textit{Varroa} infestation, or increased susceptibility. Eq. \eqref{eq:model} is modified as:
\begin{align}
    \dot{Y} &= L \frac{H+F}{w+H+F} - Y\nonumber \\ 
    \dot{O} &= (1-n)Y - H \left( \alpha - \sigma \frac{F}{H+F} \right)  \nonumber \\
    \dot{F} & = H \left( \alpha - \sigma \frac{F}{H+F} \right) - mF \, .
    \label{eq:model_breda}
\end{align}
and can be recovered in case $n=0$, where $n \in [0,1)$ is an additional mortality term on young bees, due to, e.g. \textit{Varroa} infestation or other conditions. Here, $O+Y=H$. This model has a non-zero asymptotically stable equilibrium following the same conditions for $\hat{F}$ and $\hat{H}$ as in \eqref{eq:equilibria1}, after a change $L\to L(1-n)$ and with $\hat{Y} = L \frac{\hat{H}+\hat{F}}{\hat{H}+\hat{F}+w}$ \cite{breda2022deeper}. Even in this case, there is a critical point $m_{c,3}$ at which the non-zero equilibrium vanishes and only the trivial $(0,0,0)$ equilibrium exists and is stable. Similarly to the other models, the critical parameter 
\begin{equation}
    m_{c,3} = \frac{L(1-n)}{2w\left(\alpha-\frac{L(1-n)}{w}\right)}\left(\alpha+\sigma+\sqrt{(\alpha-\sigma)^2+4\sigma\frac{L(1-n)}{w}} \right) \, 
    \label{eq:mc_breda}
\end{equation}
marks a transcritical bifurcation, whose plot is in Fig. \ref{fig:bif_diags}c.

\subsection{Quasi-static detectability of EWS}
We use the whole colony model Eq. \eqref{eq:model} to generate synthetic data and estimate EWS. The model is augmented with stochastic dynamics using the \texttt{sdeint} Python package, with the function \textit{itoint} using noise intensity of 0.01. This noise intensity is sufficiently small to produce very few noise-induced transitions near the bifurcation point. An Euler-Maruyama scheme with time step of 0.01 is used to simulate the resulting stochastic system for 10,000 days (leading to $1,000,000$ data points per simulation) around the equilibrium associated with each $m$. 

Default parameter settings are chosen as $L=2000$, $\alpha=0.25$, $\sigma=0.75$, and $w=27000$ as in the original publication \cite{khoury2011quantitative}. The control parameter $m$ is changed stepwise from 0.25 to 0.35 in steps of 0.01. The interval values correspond to reasonable $m$ values identified in \cite{khoury2011quantitative} along the model transcritical structure (see Fig.~\ref{fig:bif_diags}a), up to the critical value for $m_{c,1}$ corresponding to the chosen set of parameters. The dominant eigenvalue of the Jacobian matrix changes from -0.078 to -0.056 in this range of $m$. 

Changing $m$ in small steps over a long time horizon ensures quasi-steady state of the time series, such that EWS extraction is not confounded by commensurable timescales, and mimics a quasi-steady increase of the bee mortality rate, up to a critical value, as observed empirically \cite{russell2013dynamic}.

To investigate the effect of temporal resolution on EWS detection, we subsample the simulated data to obtain time series at different resolutions, denoted by $\rho$. We consider two time horizons: 10,000 days, corresponding to the full simulation, and 365 days. This allows us to assess how the temporal resolution and the length of the observation period affect EWS estimation.

For each combination of resolution and time series length of our synthetic data, we calculate the lag-1 autocorrelation AC(1) and variance Var, separately for each value of the 10 values of $m$:
\begin{equation}
    \mathrm{Var}_m = \frac{1}{N - 1}\sum^{N}_{i=1} (\mathcal{C}_{i,m} - \hat{\mathcal{C}}_m)^2 \ ,
\end{equation}
where $\mathcal{C}$ is the total colony size, $\mathcal{C} = H+F$, $N$ is the number of simulation points, $\hat{\mathcal{C}}$ is the mean value, and the subscript $m$ identifies values for each considered value of $m$. AC(1) is estimated in the same way.
Note that, for each $m$-value (for which we use all simulation data, as in one fixed window), the dynamics is stationary and no detrending needs to be performed. 

Next, we calculate the Kendall $\tau$ correlation of these 10 points (i.e. $\tau$ value, computed as described in Sec.~\ref{sec:EWS_from_data}, between the ten autocorrelations or variances and the numbers 1 to 10), and the corresponding p-values. The p-values are obtained from a statistical test, performed with the \texttt{scipy} function \textit{kendalltau}, which compares the observed $\tau$ value against one obtained from a null distribution, suggesting whether the value is statistically significant or likely associated with random fluctuations of the underlying data.

In summary, each datapoint in Fig.~\ref{fig:EWS-sim} represents the Kendall $\tau$ correlation calculated across ten simulated time series, each corresponding to a different value of m. For example, the datapoint at 2,000 days in the top-left panel is obtained by generating ten time series, each 2,000 days long, with $m$ taking one of ten values between 0.25 and 0.35. We calculate the lag-1 autocorrelation for each time series and then calculate Kendall's $\tau$ between these ten autocorrelation values and the corresponding ordered indices (1 to 10). The resulting $\tau$ value is plotted as a single datapoint in the figure.

\subsection{Honeybee data}
The experimental conditions are as described in \cite{frizzera2023varroa}. Two apiaries, each consisting of 5 colonies (Dadant Blatt hives: 385 ×452 ×310 mm), were established in Northern Italy. Colonies in one apiary were left untreated and were therefore naturally infested by \textit{V. destructor}, whereas colonies in the other apiary were treated with acaricides to keep mite infestation at negligible levels. Specifically, the mite population was kept under control with two strips of an amitraz-based product (Apitraz, Laboratorios Calier S.A.) applied in August, followed by three treatments with a thymol-based product in tablets (ApiLife Var, Chemical Laif S.p.A.) from September to early October. The two apiaries were set 4 km apart from each other to minimize mite transfer between them through drifting and robbing, while ensuring that the colonies in both apiaries experienced the same environmental temperatures.  

This way, all colonies with a certain property (infested or not) were grouped in the same location, within the same environment and experiencing almost equal setting. For one condition, the hives can therefore be considered (at best with the experimental controlled conditions) independent samples of one population. Hives subject to different conditions (\textit{i.e.,} infestation), on the contrary, were kept separated, but still within the same commune; this way, they remain subject to very similar setting (environment, precipitations, external temperature) but cannot cross-contaminate each other. The colonies were checked for collapse (\textit{i.e.}, number of bees in a hive equals zero) twice: first, before winter time, when several hives were already dead; second, at the following spring, to verify which colonies survived after winter. The experimental design compared colonies under infested and uninfested conditions, rather than to independently manipulate collapse as an outcome. \textit{Varroa} infestation status and subsequent colony collapse are thus strongly associated in this dataset: all uninfested colonies survived, whereas all but one infested colony (CR/12) collapsed. The analysis therefore evaluates whether temperature-derived indicators distinguish colonies under these observed outcomes, assuming that \textit{Varroa} is the driving factor underlying collapse.

Colony strength (\textit{i.e.}, the number of adult bees present in the colony) was estimated in the experimental hives approximately once a month from August to December, by counting the number of full or partial ``sixth of frames'' covered by bees in each hive at sunset and calculating the overall bee population; the conversion to the total number of bees is based on the estimation that one fully covered sixth of comb corresponds to 253 adult bees on average \cite{nazzi2012synergistic}.

Internal temperature $T_{\text{int}}$ was measured from August to November 2018 with a probe from Maxim integrated, US, having sensitivity of $\pm 0.0625^\circ$C. The probe was inserted in a queen caged held by a metal wire fasten at the top of the frame, and was positioned in the central part (c) of each hive. The sampling frequency was set to 1200 s. The central position of the probe was monitored twice a week as detailed in \cite{frizzera2023varroa}, so as to mitigate fluctuations due to displacement of the colony. Average daily external temperature data were derived from the regional meteorological observatory (ARPA FVG – OSMER and GRN, \url{https://www.meteo.fvg.it/}) to evaluate field temperature.

\subsection{Time series pre-processing}

Temperature time series must be pre-processed before extracting summary indicators. First, we consider two datasets characterized by two different sampling rates, so as to reflect realistic sampling capacities. The first dataset, $\mathcal{D}_1$, includes the mean daily temperature values used in \cite{frizzera2023varroa}; the second dataset, $\mathcal{D}_2$, uses the original data sampled every 20 minutes, as described above. $\mathcal{D}_2$ allows to extract information from data sampled as frequently as possible, but includes daily fluctuations possibly associated with fast-varying changes in colony composition, such as bees going in and out hives, as well as temperature changes within a single day, e.g., between day and night. Instead, $\mathcal{D}_1$ smooths out such fast changes, provides indication on EWS capabilities in case that high sampling rates are not possible, e.g., because of manual temperature measurements performed once a day, or different sensor setups. It also allows to make a direct comparison with the external temperature, which is provided by the local weather agency as means over one day as explained above. $\mathcal{D}_1$ and $\mathcal{D}_2$ are then pre-processed in a slightly different manner, as explained below, and provide different types of information. 
In both datasets, we distinguish between ``infested''  and ``uninfested''  colonies. Both datasets present NaN values, associated with hours or days where no sampling was available, e.g. due to sensor malfunctioning. Missing values are eliminated to avoid skewing the results, and the timestamps adjusted accordingly. The timestamps are also used to map the evolution of temperature (internal and external) and colony strength, so as to maintain a consistent cross-check between the two measures.\\

\textbf{Detrending internal temperature by external temperature.} Using $\mathcal{D}_1$, we first check whether the internal hive temperature $T^{\text{int}}$ is significantly influenced by the external temperature $T^{\text{ext}}$ and needs to be detrended by it to eliminate fluctuation due to weather conditions instead of changes in colony resilience. Rather than using correlation-based tests, we make the hypothesis that external temperature is a forcing of internal temperature and therefore causes it to change. To test the hypothesis, we use a Granger causality test. 
Let $\mathcal{H}_{<t}$ denote the history of all the variables up to time $t-1$. The quantity $\mathcal{P}(T^{\text{int}} | \mathcal{H}_{<t})$ then represents the optimal prediction of $T^{\text{int}}$ conditioned on the complete history $\mathcal{H}_{<t}$, whereas $\mathcal{H}_{<t} \backslash T^{\text{ext}}_{<t}$ denotes the same history obtained by excluding past observations of $T^{\text{ext}}$, and hence exclusively based on past observations of $T^{\text{int}}$. Then, $T^{\text{ext}}$ Granger-causes $T^{\text{int}}$ if:
\begin{equation}
    \mbox{Var}[T^{\text{int}} - \mathcal{P}(T^{\text{int}} | \mathcal{H}_{<t})] < \mbox{Var} [T^{\text{int}} - \mathcal{P}(T^{\text{int}} | \mathcal{H}_{<t} \backslash T^{\text{ext}}_{<t})].
\end{equation}

The two variances are used as forecast error and compared, so as to evaluate whether the inclusion of $T^{\text{ext}}$ improves the prediction of $T^{\text{int}}$ and thus Granger-causes it. A statistical hypothesis test complements the method, using a null hypothesis (for p-value $> 0.05$) that past values of $T^{\text{ext}}$ do not contribute to predicting $T^{\text{int}}$; if the null hypothesis is rejected, we say that one variable Granger-causes the other. The test is performed with the \textsc{Matlab} function \texttt{gctest}, checking for all possible combinations of the time series and retaining the smallest p-value over all considered lags, up to 5 days (a reasonable lag to account for thermal inertia in hives).

From Fig.~\ref{fig:Temperatures}, it is clear that the internal temperature experiences a sudden change at a certain point in time, which is around mid Autumn except for hive 8R that collapses earlier. To identify such key point rigorously, we employ the \texttt{findchangepts} function from \textsc{Matlab}, which uses a variance-based criterion to find a change point $t^*$ in time series. The Granger causality test is run on time series segments before and after the changepoint $t^*$ for both $I$ and $U$, to separate the effect of the two almost-piecewise trends and to better capture what happens before Autumn, where EWS would be particularly valuable to identify fragile colonies subjected to \textit{Varroa} stress. In all cases, except for hives CR/12 (infested) and VN (uninfested) for $t>t^*$, the null hypothesis is favored, with p-values above 0.05. For CR/12 and VN, we further test a nonlinear fit with RMSE minimization to capture the relationship between $T^{\text{ext}}$ and $T^{\text{int}}$. The best fit among all tested functional forms is anyway poor, with $R^2_{\text{adj}} \leq 0.4$ for both (see Supplementary Sec. S3.2 for details). These results suggest that detrending internal temperature by external temperature is not immediately meaningful, and that a non-parametric approach should be preferred.\\

\textbf{Detrending internal temperature with a non-parametric approach.}
As EWS are associated with quasi-stationary data, detrending the considered time series is an essential pre-processing step
\cite{dessavre2019problem}. As explained above, detrending by external temperature, or a function thereof, is uninformative. Alternatively, non-parametric methods should be preferred. Common methods include Gaussian smoothing over rolling windows, ARIMA detrending \cite{proverbio2022performance}, or LO(W)ESS smoothing \cite{bury2021deep}, particularly when seasonality is relevant. Since we do not have an \textit{a priori} reference rolling window, and since recent works have found that LOESS works best in case of time series with unknown nonlinearities \cite{o2023early}, we employ the latter. LOESS is a smoothing procedure (for univariate time series, it corresponds to LOWESS -- locally weighted scatterplot smoothing) that subtracts a smooth curve fitted by local polynomial regression from the raw time series. We apply it on an expanding window, using span 0.5. This way, we ensure that the resulted EWS are independent of seasonal trends.
For $\mathcal{D}_2$, we also observe a clear daily oscillations between day and night (see Fig.~\ref{fig:Temperatures}, bottom row, and Supplementary Sec. S4.1 for a zoom-in). We remove this additional trend by means of lag-1-day differencing. 
We identify the detrended data for internal temperature as $\mathcal{T}$.

\subsection{Calculation of summary statistics and EWS assessment from data}
\label{sec:EWS_from_data}

Each hive time series is subject to EWS assessments using a range of classical univariate indicators, for which we test the predictive capabilities and gain model-based interpretation of when they become actionable. Instead of using a rolling window method, which may be sensitive to window length \cite{proverbio2022performance}, all indicators are computed on an expanding window, which has been proposed as a valid and robust method to improve the reliability of EWS \cite{o2023early, dessavre2019problem}. Given the time series ${\mathcal{T}_1,\mathcal{T}_2,\dots,\mathcal{T}_{\text{end}}}$, the expanding window statistic at time $t$ is $\hat{\theta}(t)=f(\mathcal{T}_1,\mathcal{T}_2, \dots, \mathcal{T}_{t}),$ $t=t_0, t_0+1,\dots,T_{\text{end}}$.
The burn-in period $t_0$ mitigates initial spurious signals due to few data points in the window. We use $t_0=30$ days, that is big enough to produce reliable summary statistics and corresponds to the first summer month, usually associated with low mortality rate and low fragility of the colonies. As statistic indicators, we use the most common and significant ones, starting from the variance 
\begin{equation}
    \mathrm{Var}(t) = \frac{1}{t - 1}\sum^{t}_{i=1} (\mathcal{T}_{i} - \hat{\mathcal{T}}(t))^2
\end{equation}
computed for every time point $t$, over an expanding window including $t$ time points. $\hat{\mathcal{T}}$ is the average over the window.
We also test the lag-1 autocorrelation $\text{AC}(1) (t)$ over time and the coefficient of variation $\text{CV}(t) = \sqrt{\text{Var}(t)}/\hat{\mathcal{T}}(t)$, to remove the influence of a transcritical decreasing trend on the variance pattern. They are constructed similarly to the variance over the same expanding window. Another EWS that we test is the spectral reddening phenomenon: given the Power Spectral Density computed over an expanding window, PSD$(t)$, and its log-log slope $\beta$, we reconstruct the evolution of $\beta(t)$; in theoretical CSD setting \cite{bury2020detecting}, $\beta$ should increase from about 0 (typical of white noise) to about 2 (typical of red noise), as a sign of the ``reddening''  of the signal. Since it requires high-frequency data, spectral reddening is tested only on $\mathcal{D}_2$ and, since it is not comparable with anything from $\mathcal{D}_1$, is reported in Supplementary Sec. S4.
We also directly use the average sampled values $\hat{\mathcal{T}}(t)$: the approach to a transcritical bifurcation is a monotonous (and locally linear, close to the bifurcation point) decrease, which may be easily extracted and used as EWS. Finally, we test Min$(t)$, the minimum daily value over the considered window, which was suggested \cite{colin2025temperature} to be a promising indicator even if not linked to the phenomenon of CSD; Min is only computed on $\mathcal{D}_2$ since $\mathcal{D}_1$ already contains averages. All indicators are associated with the rightmost point of the expanding window, to perform an online analysis without including future time points.
We do not consider higher order moments such as skewness or kurtosis, which are known to be very sensitive to mean values and often return unreliable trends \cite{proverbio2023systematic}. 

We quantify the detection power of each statistical indicator using its time-varying increasing trend, which constitutes a CSD-based EWS: such increasing trends are quantified by the Kendall's $\tau$ coefficient \cite{southall2020prospects, boettiger2012early}:

\begin{equation}
    \tau = \frac{\#\text{concordant pairs} - \#\text{discordant pairs}}{t(t-1)/2} \ ,
    \label{eq:tau}
\end{equation}

where $t$ is the number of considered time points over the expanding windows, as described above. Two points $(t_1,\hat{\theta}_1)$ and $(t_2,\hat{\theta}_2)$ are a concordant pair if $\hat{\theta}_1<\hat{\theta}_2$ for $t_1<t_2$, and a discordant pair otherwise. A constant trend is expected to have $\tau=0$, an increasing trend $\tau>0$ and a perfectly monotonously increasing trend $\tau=1$.

\subsection{EWS performance}

For each colony, we evaluate whether the EWS threshold is exceeded at one or more time points during the observation period.
Treating EWS as predictor variables, this becomes a classification problem (see discussion in Supplementary Sec. S6), evaluated through sensitivity and specificity as functions of significant Kendall’s $\tau$ values from Eq. \ref{eq:tau}. To quantify performance, we use a Receiver Operating Characteristic (ROC) analysis.
For each indicator, we assess how well $\tau$ distinguishes collapsing from non-collapsing colonies, by testing whether $\tau$ exceeds a threshold $\tau_c$ at each time point. For each colony known to collapse, significant $\tau$ values before the registered collapse date (or until data end) are counted as true positives, estimating sensitivity. For each colony that does not collapse, significant $\tau$ values are counted as false positives, estimating 1-specificity.
We test all thresholds $\tau_c \in [0;1]$ and build ROC curves, which summarize the tradeoff between normalized sensitivity and specificity across thresholds \cite{zhang2015predictability}. Because multiple warning signals can be generated within the same colony over time, individual time points are not considered independent experimental replicates: the colony is treated as the independent unit of observation, while the temporal sequence of signals within each colony is used to characterize the timing and persistence of the warning. Accordingly, ROC-based performance should be interpreted as an exploratory assessment of discrimination across colonies rather than as an estimate based on independent time-point observations. We quantify discrimination using the corresponding Area Under the Curve (AUC). Due to the limited number of colonies, AUC values are interpreted descriptively rather than as independently validated estimates of predictive performance. AUC = 0.5 corresponds to random guessing, values near 1 indicate strong indicators, and AUC < 0.5 reflects worse-than-random performance. 

Our definition of ``true''  signals is intentionally conservative. Because the exact collapse date is uncertain due to low-frequency colony strength measurements, any signal before the registered collapse is accepted as a true positive, even if it may occur slightly after the actual but unobserved collapse. For the same reason, precise lead times cannot be estimated.
False signals are defined following the standard EWS interpretation: warnings of collapses that never occur. An alternative interpretation could treat EWS more generally as signals of declining resilience, in which even surviving colonies might show positive signals. However, this is less informative here, since all colonies become increasingly fragile toward winter due to the underlying transcritical dynamics. Because EWS do not quantify levels of fragility or provide precise lead times, focusing on resilience loss alone would offer limited practical information and is therefore not considered.

\subsection{Noise properties}
To obtain information about the noise influencing the empirical data, we use the full time series in a retrospective analysis. This method is only used in this case for interpretation purposes, as the extraction of EWS and their performance relies on an online approach that is blind to future data, as described above. For the analysis of noise properties, we use the high-frequency sampled dataset $\mathcal{D}_2$ and we consider data from two segments: before and after the change points $t^*$ identified before (see Tab.~\ref{tab:changepoints} for values), which identify when fluctuations start differing significantly. To remove daily and long-term trends, we first detrend the entire time series using a Savitzky-Golay (SG) filter with a window size of 35 points (half a day); also in this case, the SG filter is only employed retrospectively, as it uses future data to compute the interpolation. From the residuals, which are representative of the disturbances affecting the system, we analyze their distribution, by computing skewness $\mu$ and excess of kurtosis $\gamma^*=\gamma-3$, and their frequency properties. For the latter, we compute the Power Spectral Density (PSD) via Welch's method using \textsc{Matlab}'s \texttt{pwelch} function, with 8 segments and 50\% overlap and removing boundary artifacts. The log-log PSD-frequency ($f$) plot is fit with a log-linearized power-law, as $\log(\text{PSD}) = -\beta \cdot \log(f) + a$: the value of $\beta$ gives information about the noise color. The goodness of fit is assessed via $R^2$ statistics. We refer to Supplementary Sec. S5 for details and results.

\subsection*{Data availability}
Data are publicly available at the Zenodo repository with DOI \url{10.5281/zenodo.23339802}.

\subsection*{Code availability}
The code used for the analysis is publicly accessible at \url{https://github.com/daniele-proverbio/beeEWS} and is associated to the Zenodo DOI \url{10.5281/zenodo.23339802}.

\subsection*{Acknowledgments}
D.P. would like to thank U. Sutulovic for useful discussion.

\subsection*{Funding}
D.P. and G.G. are funded by the European Union through the ERC INSPIRE grant (project number 101076926). Views and opinions expressed are however those of the authors only and do not necessarily reflect those of the European Union or the European Research Council. Neither the European Union nor the granting authority can be held responsible for them. Temperature data were collected with funding from the European Union's Horizon 2020 research and innovation programme under grant agreement No 773921 (PoshBee) and from the Italian Ministry of Education, University and Research (MIUR), PRIN 2017 - UNICO (2017954WNT).

\subsection*{Author contributions}
Conceptualization: D.P., E.W., F.N., G.G.; Methodology: D.P., E.W.; Investigation: D.P., E.W., D.F., D.A.; Software: D.P., E.W.; Formal Analysis: D.P., E.W.; Visualization: D.P.; Data Curation: D.P., D.F.; Supervision: G.G., F.N.; Funding Acquisition: G.G., F.N.; Writing—original draft: D.P. Writing—review \& 
editing: all authors.

\subsection*{Competing interests}
The authors declare no competing interests.

\newpage
\appendix

\section*{Supplementary Information (SI)}
\section{Model-based qualitative analysis of the possible existence of tipping points}

\subsection{Qualitative stability}
The analysis of qualitative stability is particularly useful for systems where little or no quantitative information is available about the interactions among its components \cite{edelstein2005mathematical}. Qualitative stability asks whether a system is stable regardless of quantitative details such as the magnitude of the interactions, or their functional form, and is often used in ecology. It relies solely on the existence and the sign of interaction, captured within signed graphs, and on notions of monotonicity of the interactions; in this sense, it is a precursor of structural analysis \cite{BG2021structural}. To perform qualitative stability analysis, consider a signed graph such as that in Fig. 2 of the Main Text. Construct a sign matrix $A=\{a_{i,j}\}$, where entries $a_{i,j}$ are the sign of the influence (positive $+$, negative $-$, or null $0$) of element $i$ onto element $j$. Then, for a system to be qualitatively stable, the following conditions must apply \cite[Sec. 6.5]{edelstein2005mathematical}:
-\begin{enumerate}
    \item $a_{i,i} \leq 0$ for all $i$; that is, no element is positively autoregulating.
    \item $a_{i,i}<0$ for at least one $i$: at least one element is negatively self-regulating.
    \item $a_{i,j}a_{j,i}\leq0$ for all $i \neq j$. For any given pair, the elements have opposite or null effect on each other (no pairwise positive feedback).
    \item $a_{i,j}a_{j,k}\dots a_{q,r}a_{r,i}=0$ for any sequence of three or more distinct indices $i,j,\dots q,r$: there are no closed loops of interactions between three or more species.
    \item $\det A\neq 0$. All elements interact somehow with the others or with themselves.
\end{enumerate}

For the single-bee health system, whose signed graph is in Fig. 2 of the Main Text, the matrix $A$ is constructed as follows. Take $i={1, \dots, 7}$ and $j=1, \dots, 7$, associated respectively to Health, Toxic compounds, Quality of Pollen, Quality of sugar, Immunity, Pathogens, and Parasites; we do not consider the variables in circle frames from Fig. 2 nor of Quantity of Sugar, as they only provide environmental inputs without receiving additional influences by the other elements. Also, recall that Health, Toxic compounds, Pathogens and Parasites exert a negative self-regulation \cite{breda2022deeper}. Look at the sign of the interactions between elements $i$ and $j$, recalling that pointed arrows are $+$ and bullet arrows are $-$. Hence, 
\begin{equation}
    A= 
    \begin{bmatrix}
       - & - & + & + & 0 & - & - \\
       - & - & 0 & 0 & 0 & 0 & 0 \\
       0 & - & 0 & 0 & 0 & 0 & 0 \\
       0 & - & 0 & 0 & 0 & 0 & 0 \\
       + & - & 0 & 0 & 0 & - & 0 \\
       0 & 0 & 0 & 0 & - & - & 0 \\
       0 & 0 & 0 & 0 & - & 0 & - \\
    \end{bmatrix}
\end{equation}
Conditions 1 and 2 are satisfied. Condition 3, however, is already not satisfied. In fact, $a_{1,2}a_{2,1}=+$ and $a_{5,6}a_{6,5} = +$. Hence, the system is \textit{not} qualitatively stable. Note that, graphically, condition 3 corresponds to verifying that there are no pairs of like-arrows connecting a pair of elements. From Fig. 2 of Main Text we see that links \textit{k-c} and \textit{m-j} are instead both negative, contradicting the requirement. The interaction represented by link $m$ is therefore sufficient to violate the qualitative-stability condition, consistent with the structural analysis of \cite{breda2022deeper}.

\subsection{Phase plane of the dynamical model (1)}
\begin{figure}[h]
    \centering
    \includegraphics[width=\linewidth]{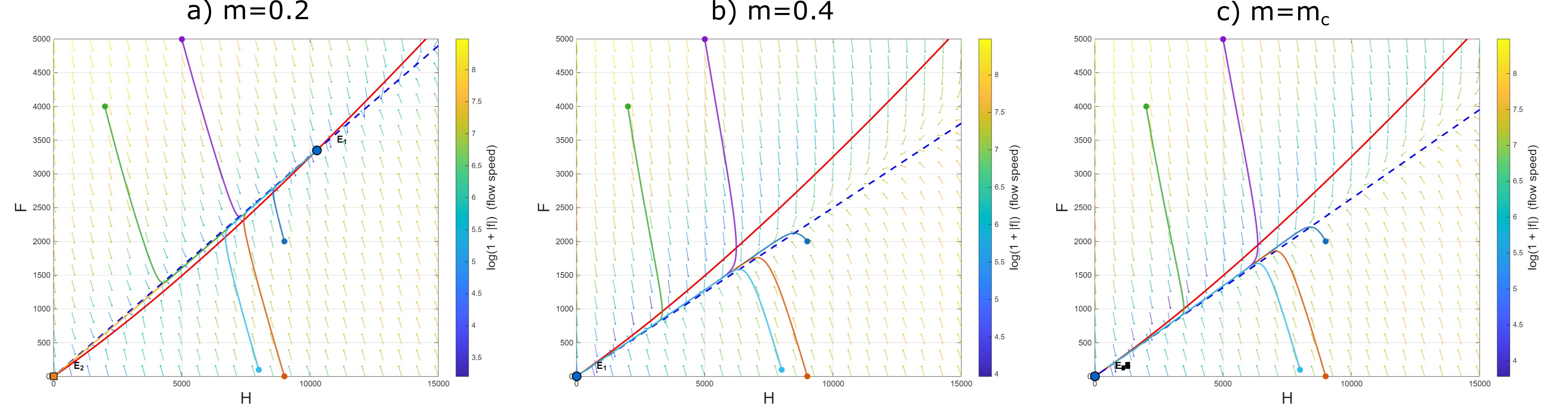}
    \caption{\small \textbf{Phase plane and vector field (color-coded according to its log-speed) of model (1) of Main Text.} Five sample trajectories are also shown, starting from the bullet points. Solid red and dashed blue lines represent the x and y nullcline, respectively. The parameters $L=2000$, $\alpha=0.25$, $\sigma=0.75$, $w=27000$ are set as in \cite{khoury2011quantitative}. The bifurcation parameter $m$ changes across panels: (a) $m=0.2$; (b) $m=0.4$; (c) $m=m_c$. Panel (a) has two stable points, panel (b) has one, while panel (c) shows a point classified as critical. Calculation performed symbolically with \textsc{Matlab}. }
    \label{fig:phase_plane}
\end{figure}

 Fig. \ref{fig:phase_plane} shows the phase plane (nullclines, vector field and equilibria) for the colony model (1) of Main Text, for three different values of the death rate parameter $m$. In panel (a), $m=0.2$: the non-zero equilibrium is stable and the zero equilibrium is unstable: randomly picked trajectories converge to an "alive" stable equilibrium. In panel (b), $m=0.4$: only the zero-equilibrium exists and is stable, and randomly picked trajectories converge to it. In panel (c), $m=m_c$: the zero-equilibrium is not hyperbolic and marks a critical point where the previous two equilibria meet and exchange stability. At point $m=m_c$, the Jacobian has a zero eigenvalue, and symbolic expansion around the equilibrium satisfies the non-degeneracy conditions for a transcritical bifurcation. 

\section{EWS using Shannon Entropy}
\begin{figure}[h!]
    \centering
    \includegraphics[width=0.9\linewidth]{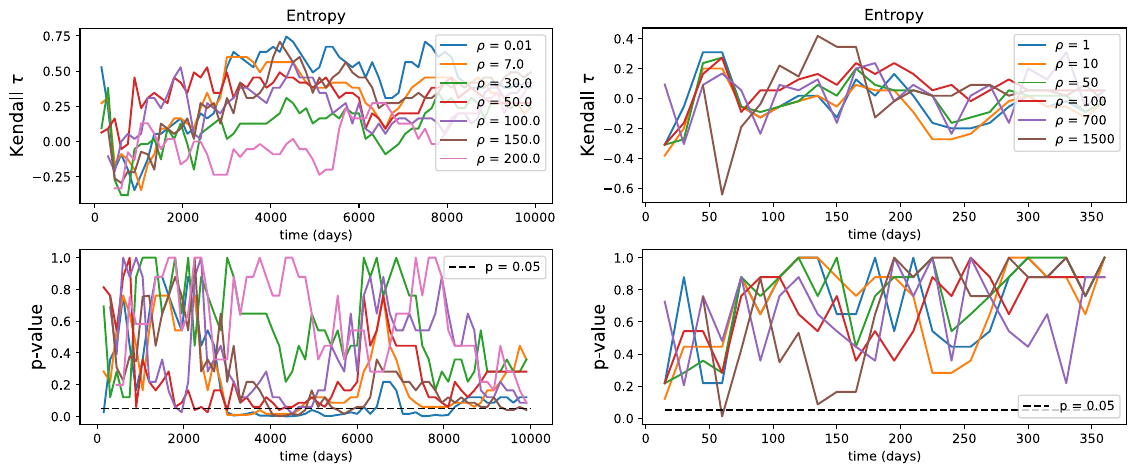}
    \caption{\small \textbf{Kendall $\tau$ and corresponding p-value extracted from trends of Shannon Entropy} ($S_H=-\frac{1}{Z}\sum_i p_i \log p_i$, where $p_i$ is the probability of a certain simulated value of colony size within the considered sample, and $Z$ is a normalization factor), calculated on simulated data of the whole colony model, as for the other indicators in Fig. 4 of the Main Text. Left column: EWS calculation using the full simulated time series; right column: using a subsample of 365 days. }
    \label{fig:entropy}
\end{figure}

Fig. \ref{fig:entropy} shows EWS and their significance estimated using Shannon Entropy. Contrary to AC(1) and Var, shown in Fig. 4 of the Main Text, EWS derived from Entropy trends do not become significant after a given time; instead, they oscillate greatly for every $\rho$, making $S_H$ an unreliable indicator at any time span. Under the white-noise conditions used in our simulations, Shannon entropy did not provide a stable EWS and was therefore not considered further. This is consistent with the findings in \cite{proverbio2023systematic}. Hence, Entropy is not further considered in the analysis.

\section{Relationship between external and internal temperature}

\subsection{Change points}
\begin{figure}
    \centering
    \includegraphics[width=0.7\linewidth]{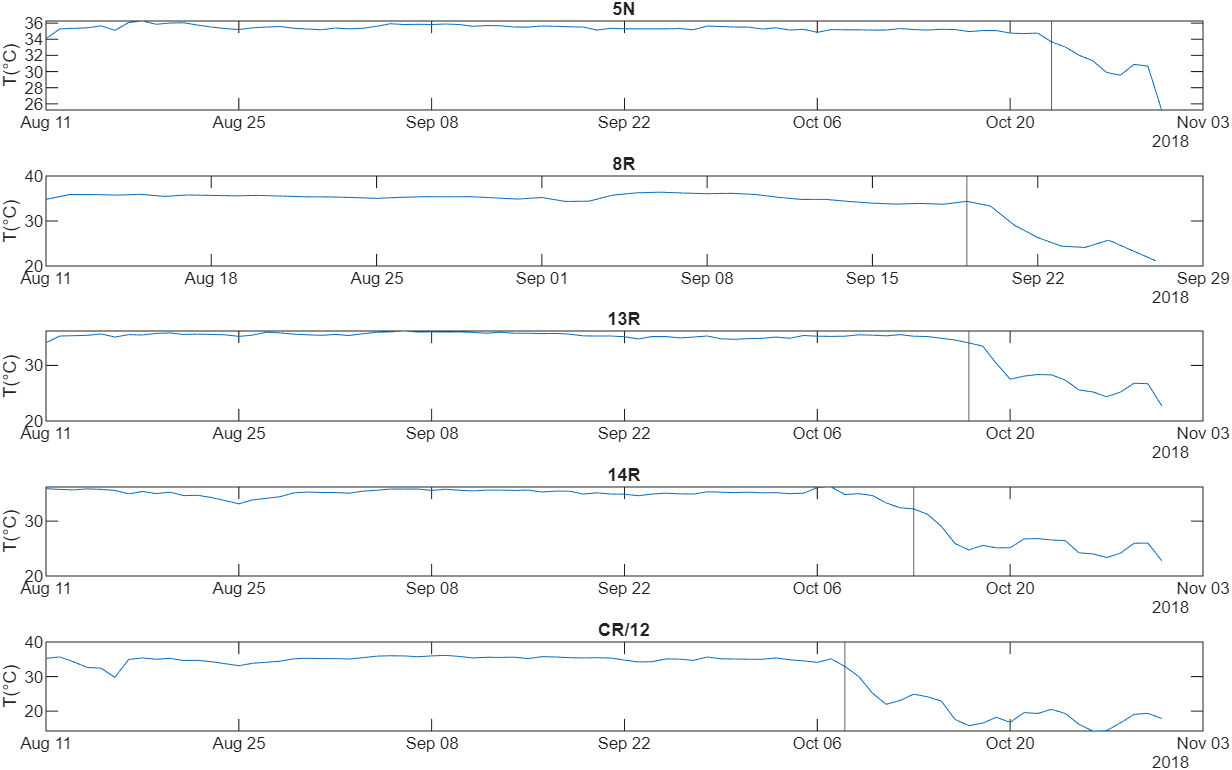}
    \caption{\small \textbf{Change point detection (vertical bar) for infested hives.}}
    \label{fig:changepoint_inf}
\end{figure}

\begin{figure}
    \centering
    \includegraphics[width=0.7\linewidth]{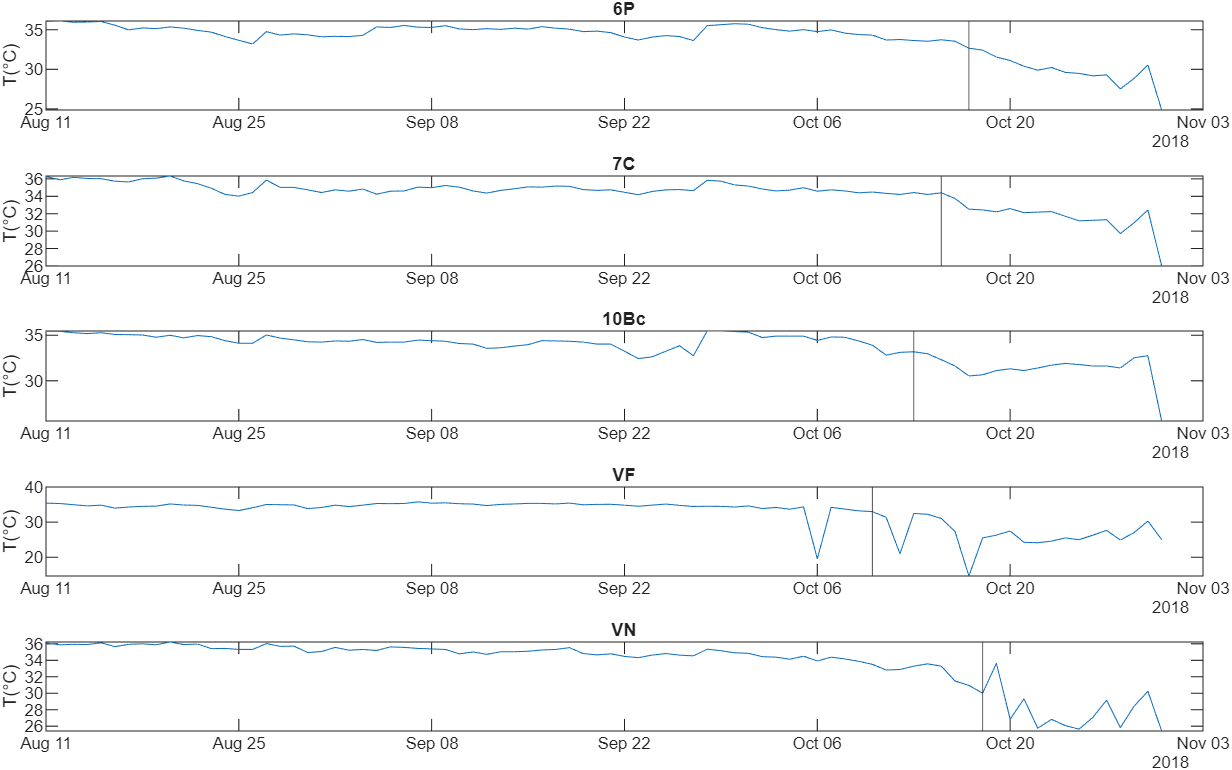}
    \caption{\small \textbf{Change point detection (vertical bar) for uninfested hives.}}
    \label{fig:changepoint_un}
\end{figure}

The points $t^*$ for each hive (change point date when internal temperature experiences a significant change in trend, identified as described in the "Time series pre-processing" section of Methods) are listed in Table \ref{tab:changepoints_1} (which duplicates Table 1 from Main Text, for ease of comparison) and visualized on the time series of each hive in Figs. \ref{fig:changepoint_inf} (for infested) and \ref{fig:changepoint_un} (for uninfested).

\begin{table}[h]
\centering
\begin{tabular}{l|lllll|lllll|}
\cline{2-11}
                             & \multicolumn{5}{c|}{Infested}                                                                    & \multicolumn{5}{c|}{Uninfested}                                                                  \\ \hline
\multicolumn{1}{|l|}{Hive} & \multicolumn{1}{l|}{5N} & \multicolumn{1}{l|}{8R} & \multicolumn{1}{l|}{13R} & \multicolumn{1}{l|}{14R} & CR/12  & \multicolumn{1}{l|}{6P} & \multicolumn{1}{l|}{7C} & \multicolumn{1}{l|}{10BC} & \multicolumn{1}{l|}{VF} & VN \\ \hline
\multicolumn{1}{|l|}{$t^*$}  & \multicolumn{1}{l|}{22/10} & \multicolumn{1}{l|}{18/09} & \multicolumn{1}{l|}{16/10} & \multicolumn{1}{l|}{12/10} & 07/10 & \multicolumn{1}{l|}{16/10} & \multicolumn{1}{l|}{14/10} & \multicolumn{1}{l|}{12/10} & \multicolumn{1}{l|}{09/10} & 17/10 \\ \hline
\end{tabular}
\caption{\textbf{Change points $t^*$ for each hive (dd/mm).} The year is 2018.}
\label{tab:changepoints_1}
\end{table}

\subsection{Relationship between external and internal temperature}

\begin{figure}
    \centering
    \includegraphics[width=0.9\linewidth]{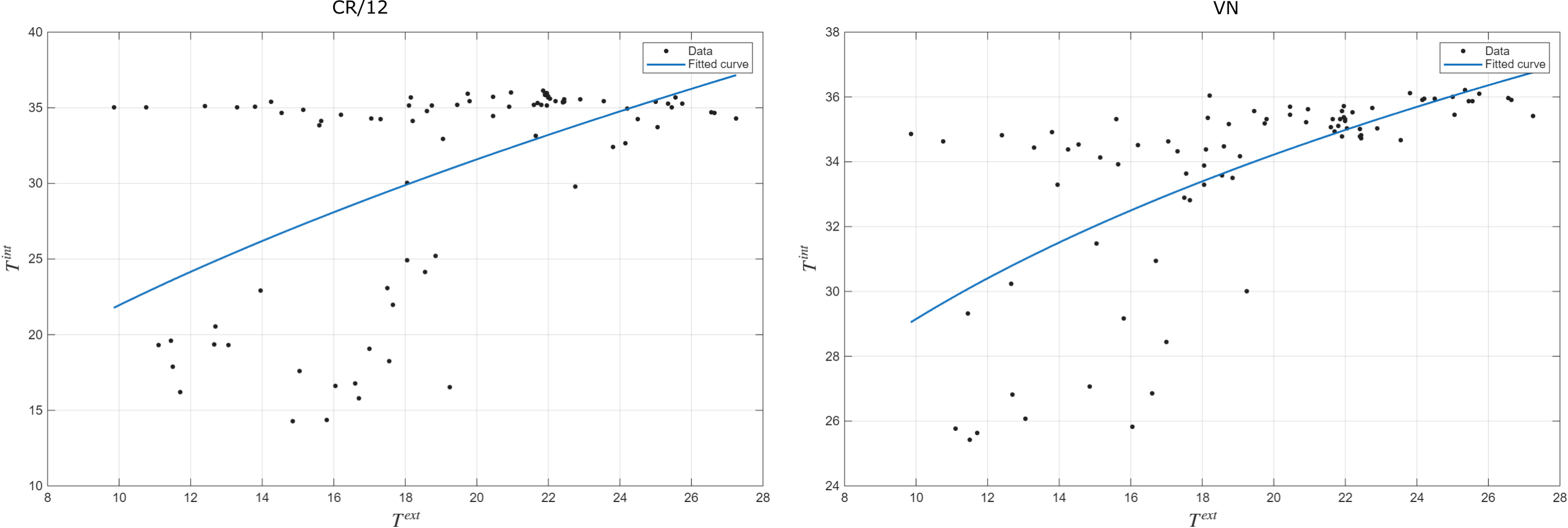}
    \caption{\small \textbf{Fit $T^{\text{int}} = f(T^{\text{ext}})$ for hive CR/12 (infested) and VN (uninfested).}}
    \label{fig:fit_temp}
\end{figure}

The absence of significant Granger causality (see Main Text) does not establish that external temperature has no physical influence on internal temperature, but indicates that its past values did not provide significant predictive information under the tested model. This result may reflect hive insulation, colony thermoregulation, differences in temporal dynamics, or other unmeasured processes.
The only case where a small but statistically significant relationship exists, is for hives CR/12 (infested) and VN (uninfested). For both, we made an attempt to fit a functional form that can describe a relationship $T^{\text{int}} = f(T^{\text{ext}})$. The best functional forms identified for both sets, with their best fit parameter values and 95\% confidence intervals, obtained using nonlinear least square minimization using \textsc{Matlab} curve fit toolbox (see Fig. \ref{fig:fit_temp}), are reported below; we recall that RMSE stands for Root Means Squared Error.

For hive CR/12, $T^{\text{int}} = a \cdot (T^{\text{ext}})^b$, with $a=6.55 \; (2.70, 10.41)$ and $b= 0.53 \; (0.33, 0.72)$. $R^2_{\text{adj}} = 0.26$, RMSE$=6.18$.

For hive VN, $T^{\text{int}} = a \cdot (T^{\text{ext}})^b$, with $a=17.11 \; (13.88, 20.34)$ and $b=0.23 \; (0.17, 0.30)$. $R^2_{\text{adj}} = 0.39$, RMSE$=2.3$.

The substantial unexplained variation (low $R^2_{\text{adj}}$) indicates that external temperature alone is insufficient to describe internal temperature dynamics and is consistent with the influence of additional environmental, colony-level, or measurement factors. Due to data unavailability, we leave the study of such drivers to future work.

\section{Indicators for $\mathcal{D}_2$}

\begin{figure}
    \centering
    \includegraphics[width=\linewidth]{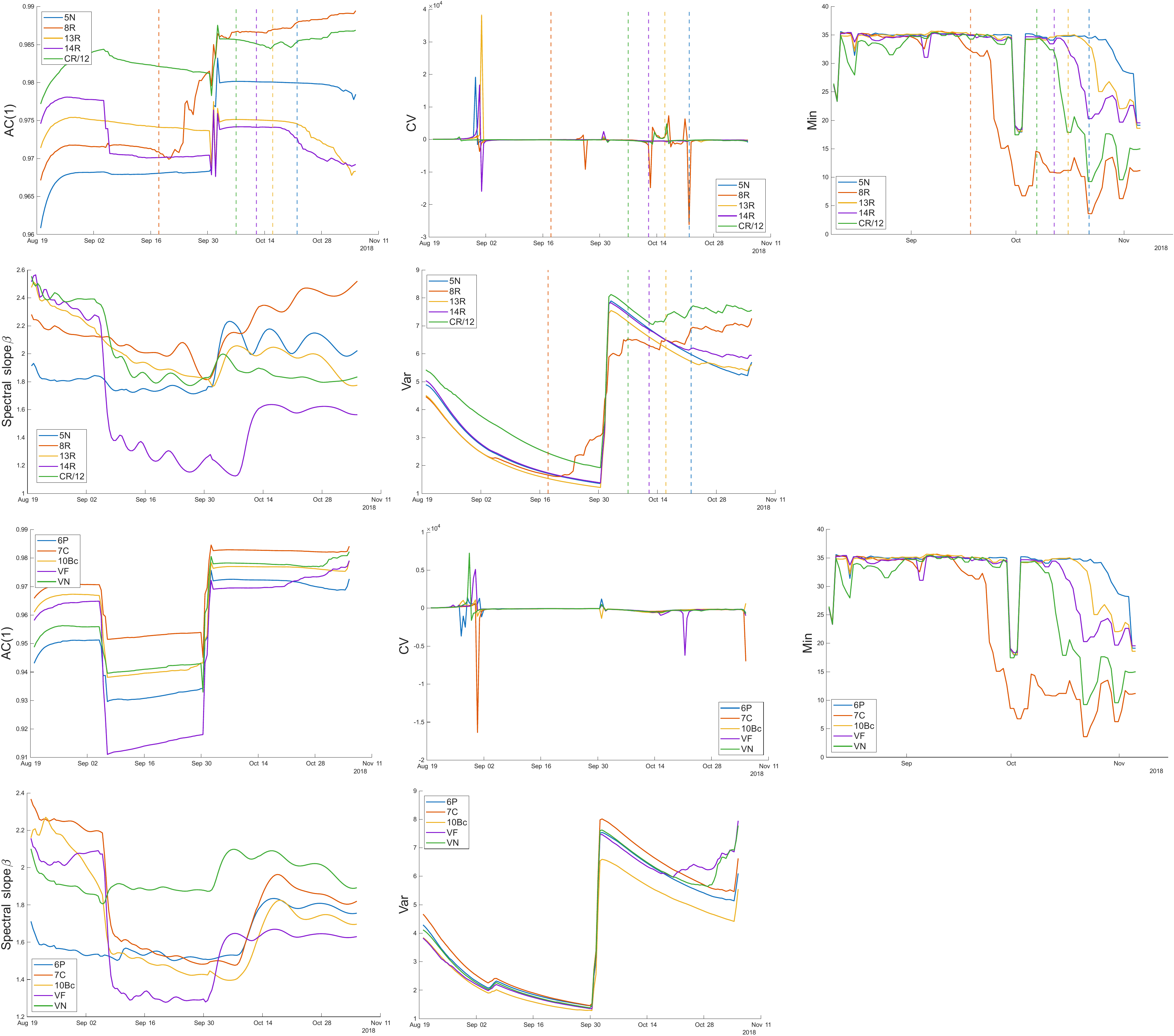}
    \caption{\small \textbf{Evolution of the considered summary indicators (CV, Var, Min, AC(1), Spectral slope $\beta$), for infested (first and second row) and uninfested (third and fourth row) hives from $\mathcal{D}_2$. }The dashed lines correspond to the identified change points $t^*$ for the infested hives, see Tab. \ref{tab:changepoints_1} for values.}
    \label{fig:indicators_d2}
\end{figure}

The evolution of CV, Var, Min and AC(1) computed from $\mathcal{D}_2$ (the high-frequency sampled dataset) are shown in Fig. \ref{fig:indicators_d2}, to complement those from $\mathcal{D}_1$ shown in the Main Text. More than in $\mathcal{D}_1$, some drastic trend changes are visible for AC(1) and Var; they coincide temporally with sensor interventions (which occurred at the end of September) or with large fluctuations in the recorded temperatures. (early September and December). These observations highlight the sensitivity of these indicators to measurement procedures and motivate caution in their interpretation.

Even by eye, Var shows trends that mostly diverge from the expected increases predicted by CSD. This fact is probably associated with the noise that is present in data sampled every 20 minutes: past works \cite{o2018stochasticity, proverbio2023systematic, kuehn2022warning} have shown that Var is sensitive to the type of noise that affects the data. Since $\mathcal{D}_2$ does not eliminate very frequent fluctuations in temperature recording, as $\mathcal{D}_1$ does (see also Sec. S5 for details), the variance may be altered and thus become not very reliable as EWS (as also reflected by the ROC analysis in Fig. 7 of the Main Text). 

The slope of the spectral density $\beta$ shows oscillations likely linked to the used algorithm. These oscillations may arise from sensitivity of the numerical PSD estimation to window length and finite-sample effects, that cannot be eliminated by simply changing the parameters of the PSD function. Because this mechanism was not independently tested here, we interpret these oscillations cautiously. Moreover, the extraction of $\beta$ relies on a numerical fit within confidence intervals (CI): it can happen that, within similar CI, the best fit $\beta$ changes slightly and looks as if it is oscillating. We also recall that PSD is estimated from relatively low-frequency and short time series, as the online extraction process uses expanding windows; hence, data may not be sufficiently highly-sampled to allow the extraction of a robust value for $\beta$. Together with the low AUC$_\beta$ value shown in the Main Text, these observations do not support the use of $\beta$ in this practical application, and demand careful additional studies for its real-world use.

\subsection{Daily oscillations in $\mathcal{D}_2$}

\begin{figure}
    \centering
    \includegraphics[width=0.7\linewidth]{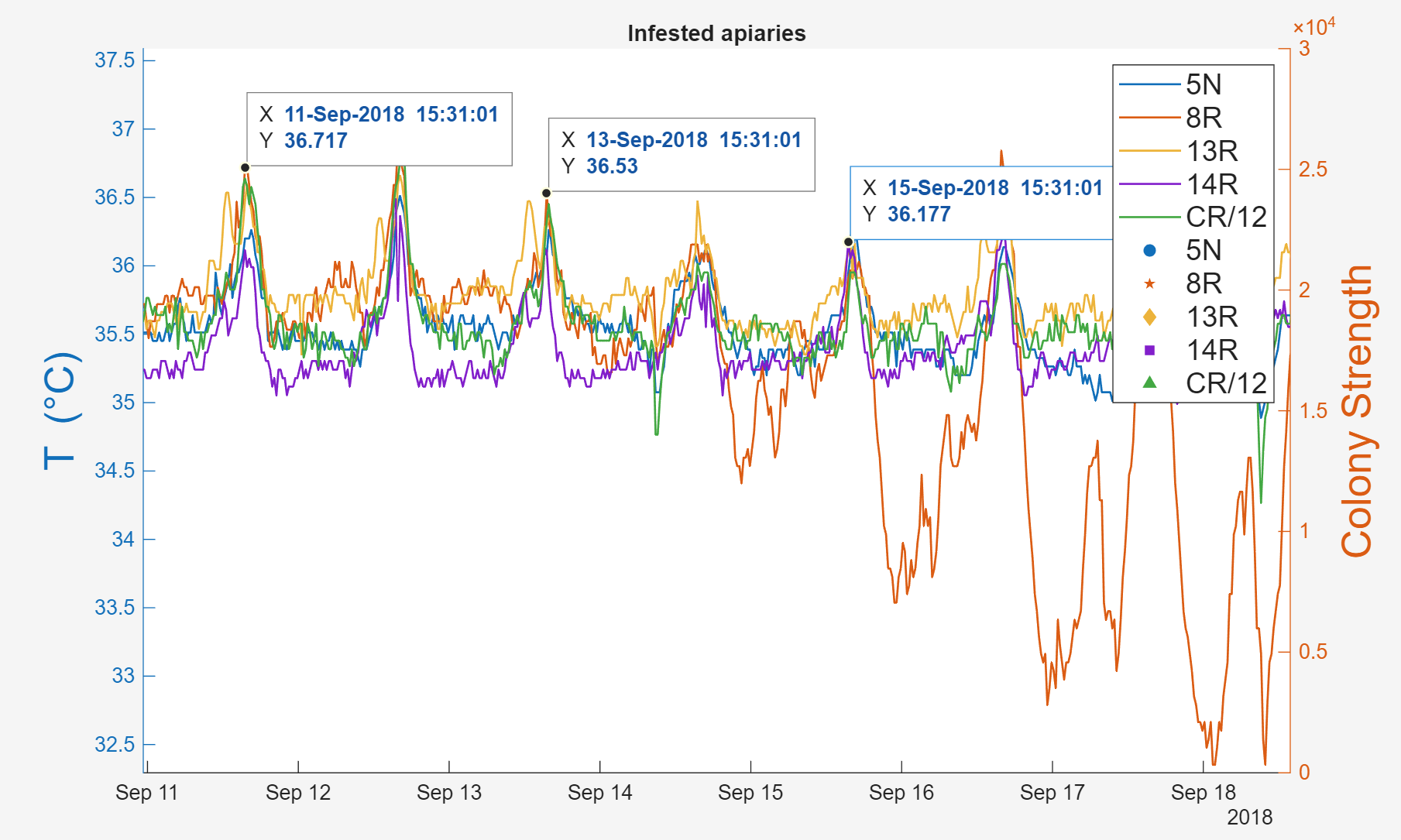}
    \caption{\small \textbf{Zoom-in into Fig. 5 (bottom-left) of Main Text}: the daily oscillations are clearly visible.}
    \label{fig:oscillations}
\end{figure}

In the Main Text, we remind that $\mathcal{D}_2$ displays clear oscillations in internal temperature between day and night. Fig. \ref{fig:oscillations} provides an example, with peaks clearly marked for consecutive days, for all infested hives.

\section{Noise properties}
\begin{figure}
    \centering
    \includegraphics[width=0.7\linewidth]{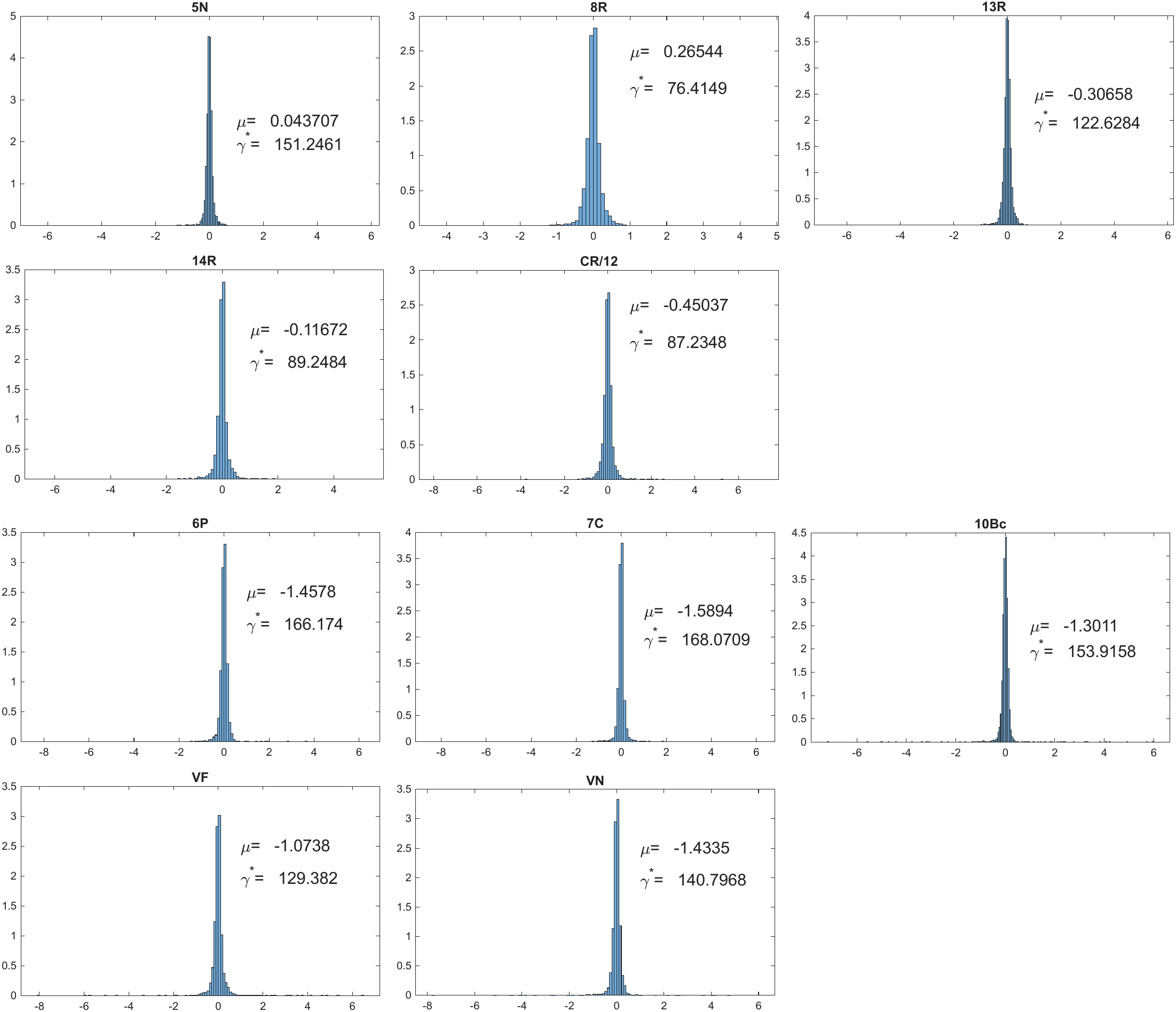}
    \caption{\small \textbf{Distribution of the detrended fluctuations for the infested (first and second row) and uninfested (third and fourth row) hives, before their respective change points $t^*$.} Skewness $\mu$ and excess of kurtosis $\gamma^*=\gamma-3$ are reported in the panels. }
    \label{fig:noise_distribution_before}
\end{figure}

\begin{figure}
    \centering
    \includegraphics[width=0.7\linewidth]{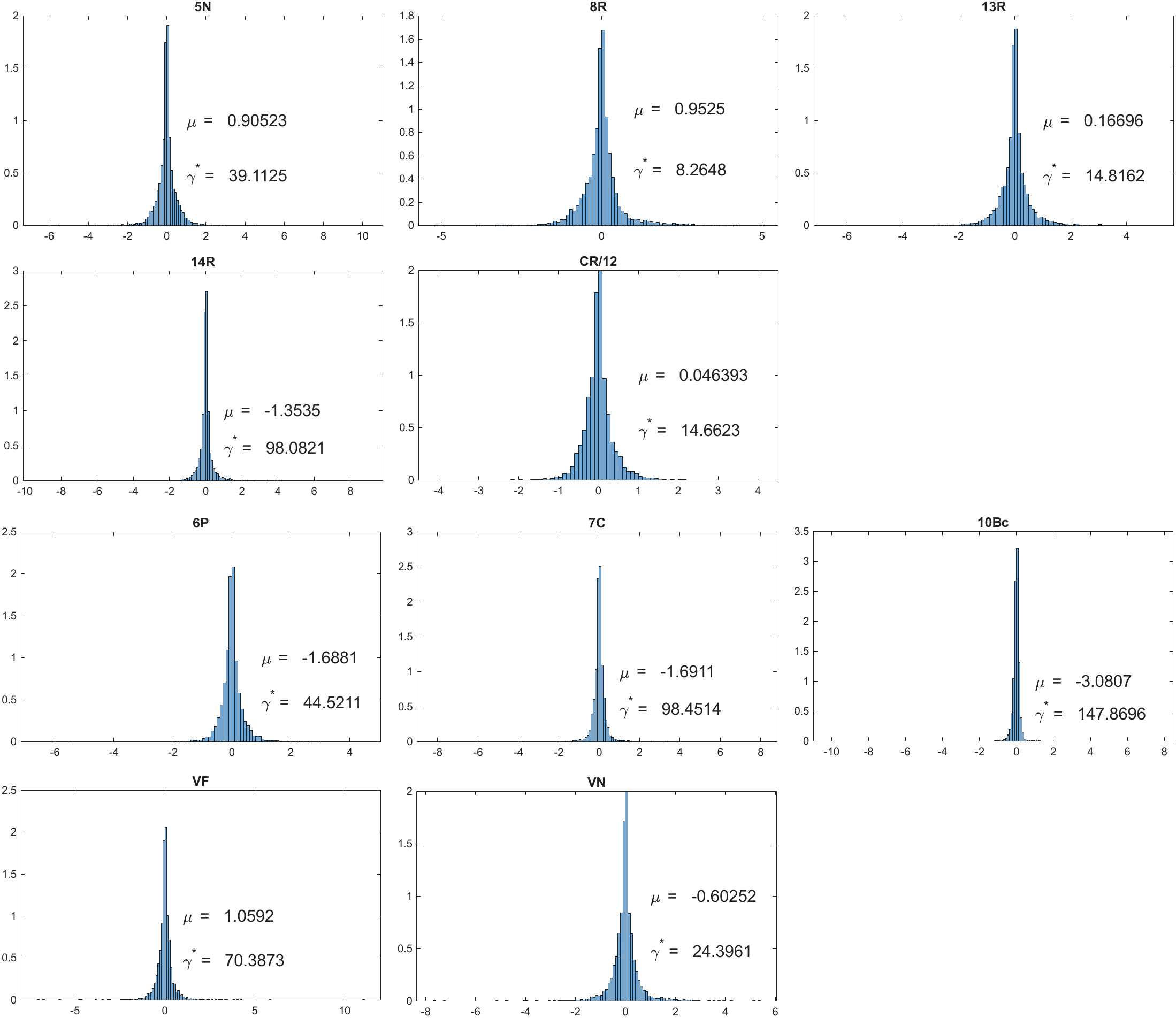}
    \caption{\small \textbf{Distribution of the detrended fluctuations for the infested (first and second row) and uninfested (third and fourth row) hives, after their respective change points $t^*$.} Skewness $\mu$ and excess of kurtosis $\gamma^*=\gamma-3$ are reported in the panels. }
    \label{fig:noise_distribution_after}
\end{figure}

Fig. \ref{fig:noise_distribution_before} shows the distribution of the detrended fluctuations (obtained as described in Methods) before the change point, together with skewness $\mu$ and excess of kurtosis $\gamma^*$. Except for hives 5N and 8R, all hives show a slightly left-skewed distribution, and all have a pronounced peak around zero. Although the magnitude of $\gamma^*$ may be emphasized by the Savitzky-Golay filtering scheme, it is nonetheless so pronounced that it is substantially different from zero. Both metrics suggest a non-Gaussian distribution. Similar conclusions can be drawn by considering the data after the change points, see Fig. \ref{fig:noise_distribution_after}.

\begin{figure}
    \centering
    \includegraphics[width=\linewidth]{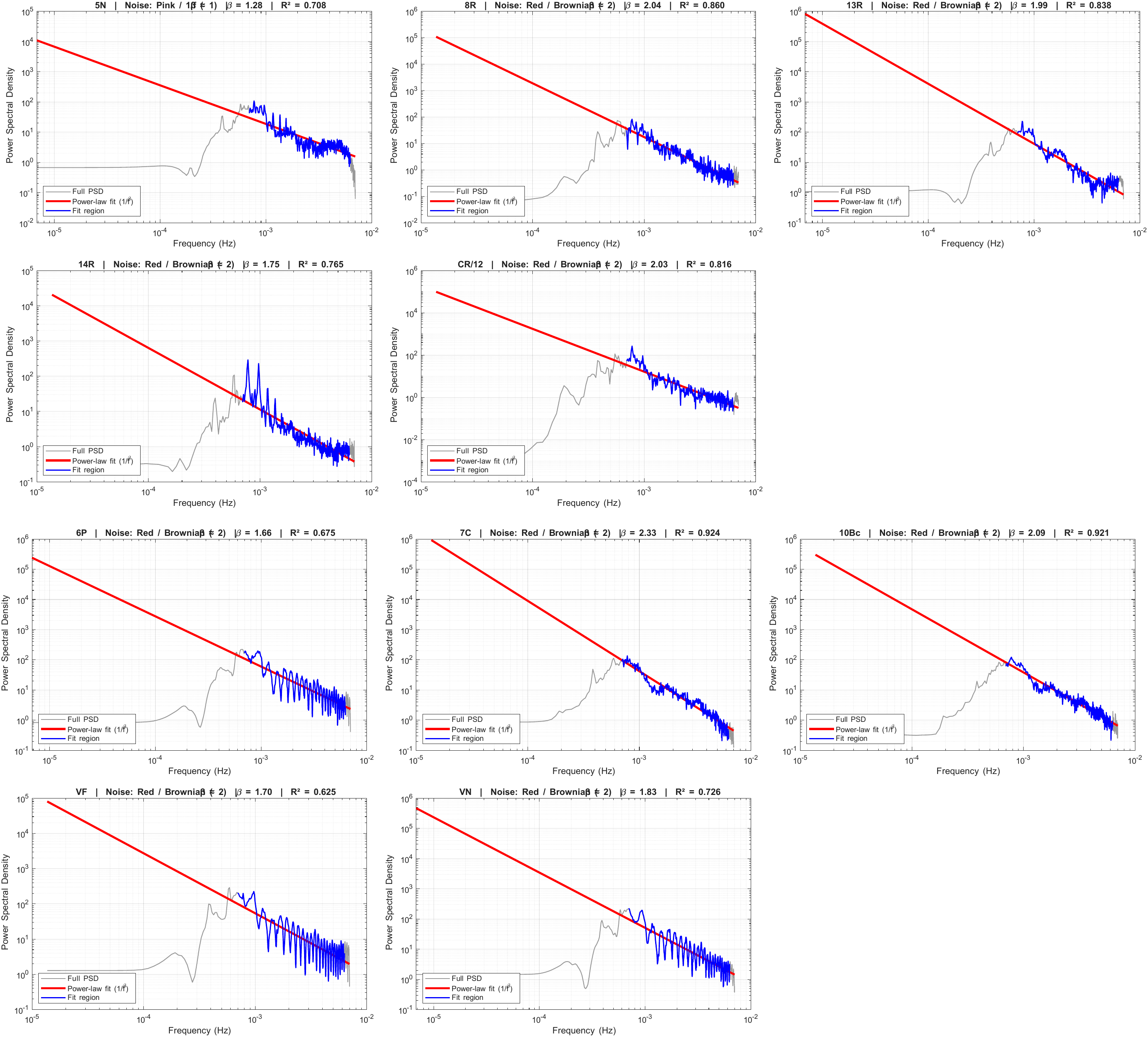}
    \caption{\small \textbf{Log-log plot of PSD vs $f$ for all considered hives, \textit{before} the change point, with linear fit $\log(\text{PSD}) = -\beta \cdot \log(f) + a$.} The value of $\beta$ and the goodness of fit $R^2$ are reported for each hive in the panels title.}
    \label{fig:noise_color_before}
\end{figure}

\begin{figure}
    \centering
    \includegraphics[width=\linewidth]{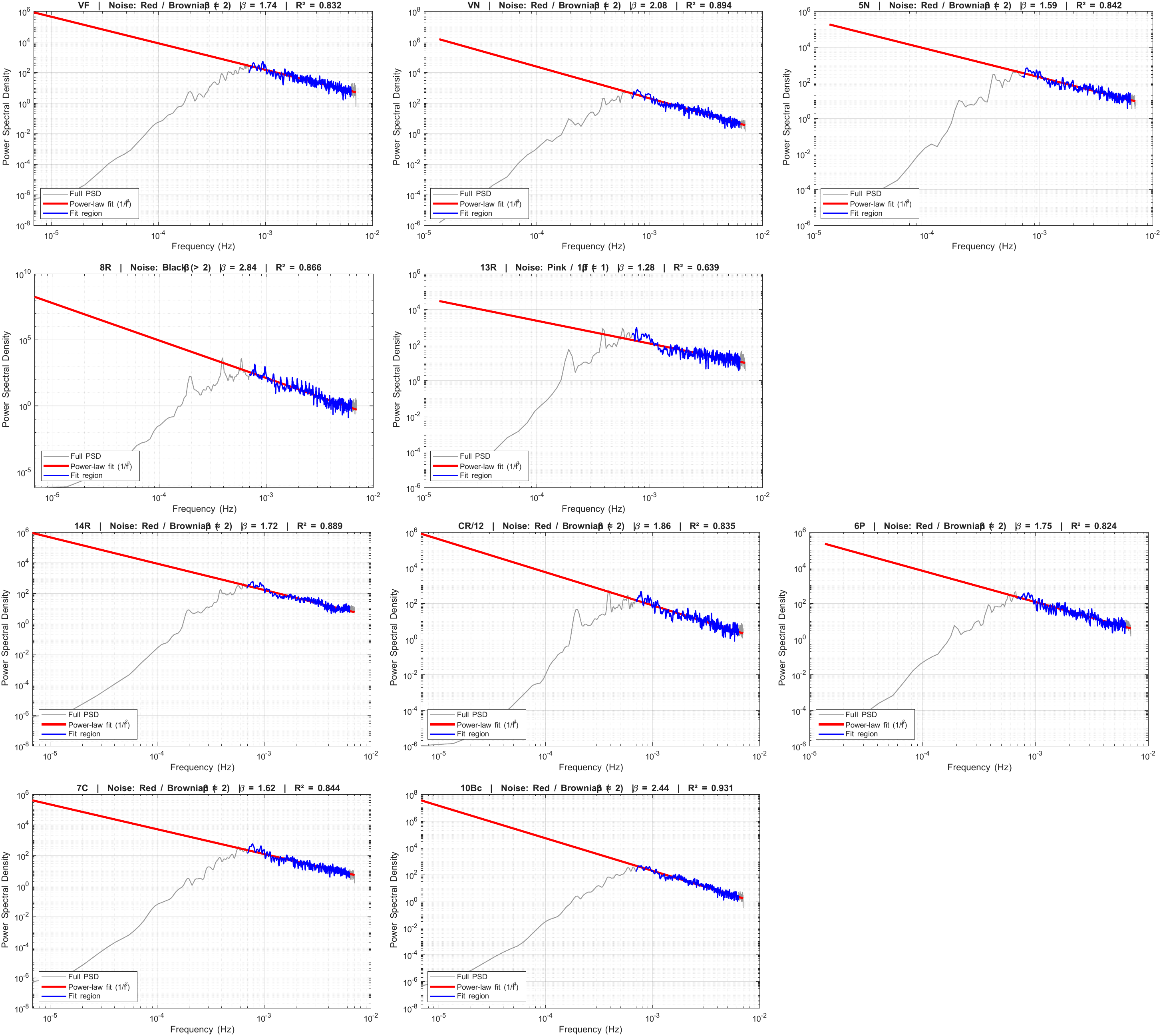}
    \caption{\small \textbf{Log-log plot of PSD vs $f$ for all considered hives, \textit{after} the change point, with linear fit $\log(\text{PSD}) = -\beta \cdot \log(f) + a$.} The value of $\beta$ and the goodness of fit $R^2$ are reported for each hive in the panels title. }
    \label{fig:noise_color_after}
\end{figure}

Fig. \ref{fig:noise_color_before} shows the PSD of all hives before change point, and the fits of its log-log slope as described in Methods. Fit statistics are given as panel titles: except for hive 5N (pink noise, $\beta \approx 1$), the fit suggests red noise ($\beta\approx 2$) in all cases. Same conclusions can be drawn by considering the data after the change point (Fig. \ref{fig:noise_color_after}), with almost all noise being classified as red.
These results provide empirical support for why idealized CSD expectations may not translate directly to hive-temperature data, as noise properties of hive-temperature records differ substantially from the idealized white-noise conditions used in the baseline simulations.

\section{Bayesian interpretation of EWS}

EWS can be viewed as predictor variables extracted from data, used to inform decisions about whether a system is approaching a regime shift. Framed this way, issuing an early alert becomes a classification problem \cite{zhang2015predictability, proverbio2022performance, bury2021deep, proverbio2024early}: distinguishing meaningful warning signals from spurious ones.  This framing highlights a distinction that is important for practical applications: the probability of observing an EWS conditional on a regime shift, $P(\text{EWS|RS})$, is not equivalent to the probability of a regime shift conditional on observing an EWS, $P(\text{RS|EWS})$. Bayes' theorem \cite{Alibakhshi2025, rietkerk2025ambiguity} gives:
\begin{equation}
    P(\text{RS}|\text{EWS}) = \frac{P(\text{EWS}|\text{RS}) \cdot P(\text{RS})}{P(\text{EWS})} \ ,
    \label{eq:bayes}
\end{equation}
where $P(\text{RS}|\text{EWS})$ is the probability of observing an RS after receiving an EWS, $P(\text{EWS}|\text{RS})$ is the probability of measuring an EWS in systems where RS exist,  $P(\text{RS})$ is the prior probability that the system can undergo a regime shift, and $P(\text{EWS})$ is the general probability of observing a signal interpreted as an EWS. 
This formulation helps address the classic "prosecutor’s fallacy" in EWS studies \cite{boettiger2012early}: confusing the probability of observing a signal before a known transition with the probability of a transition after observing a signal. Most theoretical work emphasizes the former, but practical decisions depend on the latter.
Consequently, even a theoretically expected EWS may have limited predictive value when the signal is also frequently generated by alternative processes or when regime shifts are rare. In ecological monitoring, this distinction is important because increasing variance, autocorrelation spectral reddening or other EWS may arise from environmental forcing or other non-critical dynamics.

This perspective motivates the analytical sequence adopted in this study. We first assess whether regime-shift dynamics are plausible in honeybee systems (checking if $P(\text{RS})\neq 0$) and whether the modeled transition generates CSD-based signals (checking if $P(\text{EWS|RS}) > 0$). Quantifying those probabilities more precisely is left to future work. Our approach, based on structural tests \cite{PKG2025}, presents an alternative to studies such as \cite{o2023early}, aimed at testing from empirical observations whether a system had been subject to critical transitions in the past, or to model-based analysis of tipping elements \cite{rietkerk2025ambiguity}, performed when mechanistic models exist. 
We then examine whether these signals can be detected under realistic sampling and noise conditions and, finally, whether their observation discriminates colonies that subsequently collapse from those that survive. Thus, theoretical expectation, empirical detectability and predictive utility represent distinct levels of EWS assessment.


Overall, the Bayesian framing helps clarify that EWS should not be interpreted as universal predictors, but as probabilistic evidence whose usefulness depends on system dynamics, alternative explanations for signals, and the baseline likelihood of tipping itself. This perspective can improve both the interpretation and practical deployment of early warning signals in ecology and beyond.

\end{document}